\documentclass[12pt]{article}
\usepackage{epsfig}

\newcommand{\pfrac}[2]{\left(\frac{#1}{#2}\right)}
\newcommand{\GeV}{{\rm\,GeV}}
\newcommand{\MeV}{{\rm\,MeV}}
\newcommand{\be}{\begin{equation}}
\newcommand{\ee}{\end{equation}}

\begin{document}
\thispagestyle{empty}
\begin{flushright}
MZ-TH/01-15\\
hep-ph/0105227\\
May 2001
\end{flushright}
\vspace{0.1cm}
\begin{center}
{\Large\bf Spectral moments of two-point correlators\\[7pt]
in perturbation theory and beyond}\\[16pt]
{\large S.~Groote$^1$, J.G.~K\"orner$^1$ and A.A.~Pivovarov$^{1,2}$}\\[12pt]
$^1$ Institut f\"ur Physik der Johannes-Gutenberg-Universit\"at,\\[3pt]
Staudinger Weg 7, 55099 Mainz, Germany\\[7pt]
$^2$ Institute for Nuclear Research of the\\[3pt]
  Russian Academy of Sciences, Moscow 117312
\end{center}
\vspace{0.5cm}
\begin{abstract}\noindent
We discuss the choice of weight functions for the moments of the spectral
density for two-point correlators of hadronic currents over a finite energy
interval. Of phenomenological relevance is an analysis of the spectra of
$\tau$ lepton decay on the energy interval $[0,M_\tau]$ and the low-energy
hadron production in $e^+e^-$ annihilation. General arguments are given for
the calculability of such moments in perturbation theory using both a
finite-order analysis and infinite resummation on the contour. Nonperturbative
contributions emerging from the operator product expansion for two-point
correlators are discussed within explicit models for the physical spectra. The
quantitative analysis strongly disfavours weight functions that suppress the
high-energy contribution to theoretical moments. This is in agreement with
expectations from qualitative considerations in perturbative QCD with
asymptotic freedom. We discuss the implication of our results for the ultimate
accuracy that can be reached in $\tau$ decays and low-energy $e^+e^-$
annihilation into hadrons with present experimental data.
\end{abstract}

\vspace{3pt}

\vspace{3pt}

\newpage

\section{Introduction}
The central quantities of interest for studying strong interactions in
low-energy hadron phenomenology at accelerators are the spectra of produced
particles as being related to the cross section and described by spectral
functions. The modern theory of strong interactions (QCD) describes the
production processes within perturbative expansions in the strong coupling
constant in terms of quarks and gluons and is unable to predict the low-energy
hadron spectra pointwise as functions of the energy. Only some integral
characteristics of low-energy hadron spectra such as the average over an
energy interval can be reliably calculated within perturbative QCD.
Theoretically, the procedure of averaging is best justified for observables
related to two-point correlators of hadronic currents with well established
and simple analytic properties as functions of the energy. Phenomenologically,
this is the case for $e^+e^-$ annihilation into hadrons and hadronic $\tau$
lepton decays where experimental data of high precision are available.

The process of $e^+e^-$ annihilation into hadrons has been a laboratory for
studying strong interactions since long ago where important results have been
obtained (as a recent review, see e.g.\ Ref.~\cite{eeann}). As a famous
example of such results one can mention the constraint on the number of
colours $N_c$ in QCD obtained from the normalized decay rate $R_{e^+e^-}(s)$
of $e^+e^-$ annihilation into hadrons smoothed over a finite interval at low
energies of about $\sqrt s\approx 2\div 3\GeV$
\[
N_c=\frac32R_{e^+e^-}(s)\left(1+O(\alpha_s)\right)\approx 3
\]
where $\alpha_s$ is a strong coupling constant. During the last decade much
attention has also been paid to a study of hadronic (semileptonic) $\tau$
lepton decays which provides a wealth of information on low-energy hadronic
physics where the accuracy of experimental data is permanently
improving~\cite{exp,PDG}. This makes $\tau$ lepton physics an important area
of high-precision particle phenomenology.

In the present paper we discuss in some detail the choice of the moments which
are given by integrals of the spectral density over the energy range available
in experiments. These integrals may contain specific weight functions of
different shape that determine the particular form of averaging and the kind
of moments. The moments with suitable weight functions can be computed
theoretically in perturbative QCD while the detailed pointwise description of
low-energy spectral functions themselves as functions of the energy requires a
quantitative understanding of the hadronization mechanism which is lacking at
present, especially for light hadrons. The procedure of averaging, in its
different modifications, is referred to as a property of duality between
hadronic and quark-gluon spectra~\cite{dual}.

While generating a particular set of the moments of the spectral density as
observables to confronting theory with experiment, one faces the usual
conflict between the precision requirements dictated by theory and experiment.
In perturbative QCD a reliable computation is possible only at large energies
where the perturbation series converges in a sense that a finite number of
high-order corrections in the strong coupling constant (usually one or two) is
considered to be numerically small because of the property of asymptotic
freedom, so that the moments which emphasize the high-energy part of the
spectrum are most reliably calculated. However, such moments are not welcome
from the experimental point of view because contributions coming from the
high-energy part of the spectrum in the range $0\le s\le M_\tau^2$ usually
have got a poor experimental accuracy. On the other hand, moments which
suppress the high-energy part of the spectrum are saturated by low-lying
resonances or few-particle states which are easier to measure experimentally
compared to the states with many particles. This gives rise to better
experimental precision for such moments, which would induce on using these
moments for a theoretical analysis. However, the theoretical uncertainty for
such moments is larger because of both the poor convergence of the
perturbation series due to the large value of the effective coupling and the
large nonperturbative contributions stemming from the infrared region which
are difficult to estimate quantitatively.

The semiphenomenological way for estimating the nonperturbative contributions
based on the operator product expansion (OPE) for two-point correlators is
most advanced where the knowledge of the numerical values for a few
low-dimension terms (gluon condensate and four-quark condensates) fixed from
the data. Higher order terms of dimensionality larger than six are virtually
unknown; the numerical values that are commonly used derive from some
approximations (mainly the factorization into condensates of lower
dimensionality) or models (instantons for gluonic condensates) with
theoretical uncertainties being scattered over a wide range with rather
hypothetical justification in the literature. This makes the estimate of
contributions of nonperturbative terms of higher orders rather imprecise.

Nowadays it is a general trend to consider restrictions imposed by 
perturbation theory convergence to become more important than those coming
from nonperturbative terms. With the knowledge of higher and higher order
terms of the perturbation series the asymptotic character of perturbation
theory is more and more pronounced which forces one to move the
renormalization point to larger and larger energies. In this region
nonperturbative terms die out fast because being power corrections they
decrease faster than the terms of the logarithmic perturbation series. On the
other hand, power corrections reveal the general structure of observables
computed in QCD, i.e.\ they account for nonperturbative effects. In this
circumstances there is a tendency to introduce power corrections almost by
brute force in case where OPE is not directly applicable because their
existence seems to be justified by general principles.  Even though the
motivation for such a development is clear, the present state of the analysis
is definitely far from being quantitative. The obstacle is also that this kind
of analysis is infrared dependent, i.e.\ it depends on the properties of
strong interactions at small energies where perturbation theory -- the only
universal tool of quantitative investigation -- is not applicable. This fact
makes conclusions based on power corrections depend on the particular model or
convention (for instance on a recipe of resummation).

The main purpose of this paper is to give explicit arguments and
quantitative estimates for the calculability of the moments of two-point
correlators in perturbative QCD and to find a compromise choice of weight
functions for the moments that optimizes the precision comparison of
theoretical input with experimental data. To be specific, we are talking about
the precision analysis of $\tau$ lepton decays having been of much interest
recently where experimental data are very good. Therefore, we concentrate on
the aspects of accuracy of theoretical calculations for the QCD part of the
differential $\tau$ lepton decay rate and its moments. It is the high
precision achieved in the experimental analysis of $\tau$ decays and the
rather advanced stage of theoretical description that calls for a critical
examination of the theoretical formulae used in the analysis of the physics of
the $\tau$ system. In particular the criteria of reliability of theoretical
formulae should be considered carefully. The quantitative discussion of this
issue is possible because of the simplicity of the Green's functions necessary
to describe the process -- only two-point correlators depending on one single
energy variable are relevant. The analytic properties of the two-point
correlators  in the energy variable are strictly fixed from general principles
of quantum field theory. An additional simplification for the renormalization
group analysis is provided by the fact that only one scale is involved in the
theoretical description. 

Theoretical calculations for the $\tau$ system can be done within OPE for
two-point correlators which contain perturbative and power corrections.
Perturbation theory can then be further improved using methods of summation of
an infinite number of terms within the renormalization group. We consider
these steps in turn. In Sec.~2 we give the basic ideas and fix the notation.
In Sec.~3 we consider finite-order perturbation theory, in Sec.~4 we consider
power corrections. As an important phenomenological example we consider some
peculiarities related to quark mass corrections to current correlators which
are important for the determination of the strange quark mass (Sec.~5). In
Sec.~6 we consider the infinite resummation of perturbative terms generated by
the running of the coupling constant within a renormalization group approach.
Sec.~7 contains our conclusions. 

\section{Setting the stage:
  details of the theoretical description}
Having described the motivation of our work which is rather general and valid
for two-point correlators of hadronic currents, we now give the specific
details for the description of $\tau$ decay observables. Similarities to the
$e^+e^-$ annihilation into hadrons are obvious and applications to other
hadronic channels are straightforward. In this section we mainly fix our
notation; the consideration is rather standard and can be found in the
numerous literature on this subject~\cite{pichrev,Chetyrkin:1998ej}.

The semileptonic (hadronic) $\tau$ lepton decay is mediated by the charged
weak hadronic current of the form
\begin{equation}\label{weak}
j^w_\mu(x)=V_{ud}\bar u\gamma_\mu(1-\gamma_5)d
  +V_{us}\bar u\gamma_\mu(1-\gamma_5)s
\end{equation}
where $V_{ud}$ and $V_{us}$ are Cabibbo-Kobayashi-Maskawa matrix elements
(elements of the weak mixing matrix). The correlator for the weak hadronic
currents in
Eq.~(\ref{weak}) has the general form
\begin{equation}\label{correlator}
\Pi_{\mu\nu}(q^2)=12\pi^2 i\int\langle Tj_\mu(x)j_\nu^\dagger(0)\rangle
  e^{iqx}dx=q_\mu q_\nu\Pi_q(q^2)+g_{\mu\nu}\Pi_g(q^2)
\end{equation}
where $\Pi_q(q^2)$ and $\Pi_g(q^2)$ are invariant scalar functions. These
scalar functions are further specified depending on which current is
considered. In case of the $(\bar ud)$ quark current
$j_\mu(x)=\bar u\gamma_\mu(1-\gamma_5)d$ (denoted as light quark case) the
massless limit is assumed in which case the correlator is transverse, i.e.\
both invariant functions $\Pi_{q,g}(q^2)$ are expressible through a single
scalar correlator function $\Pi_{ud}(q^2)$
\begin{equation}\label{scalfun}
\Pi_q(q^2)=\Pi_{ud}(q^2),\qquad\Pi_g(q^2)=-q^2\Pi_{ud}(q^2).
\end{equation}
The correlator in case of the $(\bar us)$ quark current (the term proportional
to $V_{us}$, also referred to as the strange quark case) is slightly different
as the nonvanishing strange quark mass is taken into account. The relevant
formulae are given e.g.\ in
Refs.~\cite{Chetyrkin:1998ej,Korner:2000wd,pichprades}. We will discuss the
strange case later. The rest of the consideration is rather similar to the
light quark case. We skip the specification $(\bar ud)$ in the following and
use $\Pi_{ud}(q^2)\equiv\Pi(q^2)$. We mention that we work within QCD with
three light quarks and do not consider corrections due to heavy quarks
($c$ quark) which would enter the calculation at high orders of perturbation
theory through loop effects~\cite{chet93,mclar,pivmccorr}. In addition to the
different current specifications as in Eqs.~(\ref{weak}) and~(\ref{scalfun})
it is convenient to consider the vector and axial parts of the correlator
separately,
\begin{equation}\label{correlatorsep}
\Pi_{V+A}(q^2)=\Pi_V(q^2)+\Pi_A(q^2)
\end{equation}
with $\Pi_V(q^2)$ beging related to the pure vector part and $\Pi_A(q^2)$
related to the pure axial part. We introduce the spectral density of a
correlator as the discontinuity across the physical cut,
\begin{equation}
\rho(s)=\frac1{2\pi i}{\rm Disc\,}\Pi(s)
  =\frac1{2\pi i}(\Pi(s+i0)-\Pi(s-i0)),\qquad s>0
\end{equation}
which splits into vector and axial-vector parts accordingly,
\begin{equation}\label{correlatorsep1}
\rho_{V+A}(s)=\rho_V(s)+\rho_A(s).
\end{equation}
Here we mainly concentrate on the massless limit for the correlator. The case
of the strange quark with nonvanishing mass will be discussed later. In the
following the indices $V$ and $A$ will be omitted in the generic case as well
if no confusion arises. The correlator in Eq.~(\ref{correlator}) is
normalized to the number of colours $N_c$ (the same holds true for the spectral
density which means that $\rho(s)\rightarrow N_c$ for $s\rightarrow\infty$) in
the leading parton model approximation with massless quarks. In the following
we will occasionally use also a slightly different normalization which
explicitly accounts for the number of colours, resulting in a correlator which
is normalized to unity.

The above considerations are general and are also used by experimentalists to
classify the appropiate channels: strange particles ($K$ mesons) form the
strange channel, non-strange axial-vector mesons (as $a_1$ and the like) 
form the axial-vector channel, and the classical vector meson $\rho$
represents the non-strange vector channel. In this respect the pion is
somewhat special. It is a Goldstone boson with spin zero and gives a
contribution to the axial correlator $\Pi_A(q^2)$ in the massless limit. 

Before specifying the theoretical calculations we give the general form of an
important observable of $\tau$ decays. The basic observable is the normalized
$\tau$ lepton decay rate for the decay of the $\tau$ lepton into hadrons
written in the standard form as
\begin{equation}\label{rate}
R_\tau=\frac{\Gamma(\tau\rightarrow{\rm hadrons}
  +\nu_\tau)}{\Gamma(\tau\rightarrow l+\bar\nu_l+\nu_\tau)}=N_cS_{\rm EW}
  \left(|V_{ud}|^2(1+\delta_{ud})+|V_{us}|^2(1+\delta_{us})\right).
\end{equation}
The leading terms in Eq.~(\ref{rate}) are the parton model results while the
terms $\delta_{ud}$ and $\delta_{us}$ represent the effects of QCD
interactions and mass effects (in case of nonvanishing quark
masses)~\cite{SchTra84,Bra88,Bra89,NarPic88,BraNarPic92}. $V_{ud}$ and
$V_{us}$ are matrix elements of the weak mixing matrix as defined in
Eq.~(\ref{weak}), and $S_{\rm EW}$ describes the electroweak radiative
corrections to the $\tau$ decay rate~\cite{braw}. Now we take a deeper look
into the physics of hadronic $\tau$ decays. In the massless limit the
expression for the $\tau$ decay rate relevant for the QCD part of the quark
current is given by the phase-space integral
\begin{equation}\label{int}
R_\tau^{\rm QCD}=N_c\int_0^{M_\tau^2}\left(1-\frac{s}{M_\tau^2}\right)^2
  \left(1+2\frac{s}{M_\tau^2}\right)\rho_{V+A}(s)\frac{ds}{M_\tau^2}.
\end{equation}
The spectral density $\rho(s)$ is related to Adler's function $D(Q^2)$ through
the dispersion relation
\begin{equation}\label{disp}
D(Q^2)=-Q^2\frac{d}{dQ^2}\Pi(Q^2)=Q^2\int\frac{\rho(s)ds}{(s+Q^2)^2} 
\end{equation}
where $Q^2=-q^2$. At this point theory enters. It is Adler's function $D(Q^2)$
that is most convenient to compute theoretically in the Euclidean domain. It
allows one to theoretically predict $\tau$ observables through a theoretically
computable spectral density $\rho(s)$. Still, to extract the theoretical
prediction for $\rho(s)$ from $D(Q^2)$ is not straightforward. The point is
that $D(Q^2)$ is only known as an asymptotic expansion at large Euclidean
values $Q^2$ while $\rho(s)$ is obtained as a discontinuity of the correlator
across the physical cut. Therefore, an analytic continuation into the complex
$q^2$-plane is necessary. Analytic continuation is an improperly posed
problem, i.e.\ small errors in the initial function $D(Q^2)$ can produce big
errors in $\rho(s)$. This instability is especially important for a theoretical
computation of $\rho(s)$ at low energies. This situation can also be
reformulated in the language of integral equations. The dispersion relation in
Eq.~(\ref{disp}) gives the spectral density $\rho(s)$ through Adler's function
$D(Q^2)$ as a solution of the integral equation. The integral equation given
by Eq.~(\ref{disp}) is a Fredholm integral equation of the second kind which
is known to lead to an improperly posed problem. Thus, errors for the function
$\rho(s)$ (as solution of this integral equation) are not continuously related
to errors for $D(Q^2)$ (as initial data of the integral equation) and can be
very large. The general procedure of constructing approximate solutions for
such a problem was suggested by Tikhonov and is known as smoothing. Averaging
the spectral density over a finite energy interval (as in sum rules that
correspond to the duality concept) can be considered as a particular
realization of Tikhonov's smoothing procedure.

\begin{figure}
\epsfig{figure=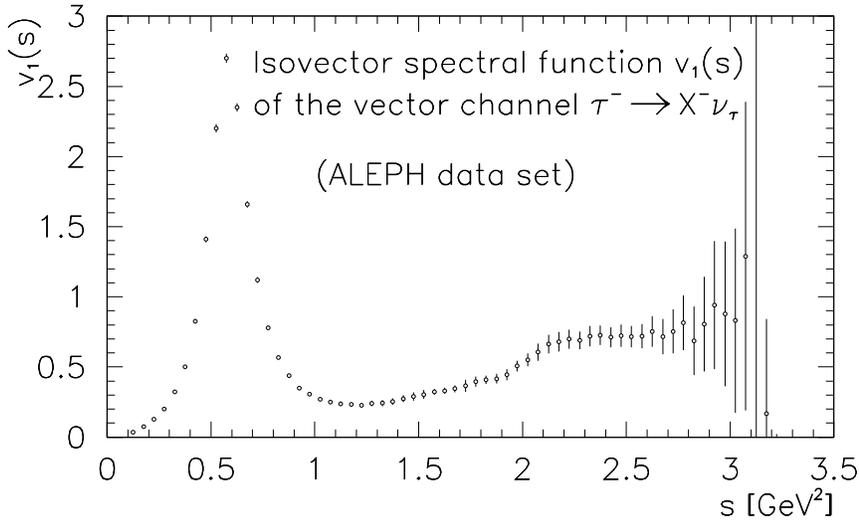, scale=0.7}
\caption{\label{fig1}experimentally measured values for the relative cross
section in the $e^+e^-$ channel, the data set is taken from
Ref.~\cite{Aleph97}.}
\end{figure}

The function $\rho(s)$ at low energies is the main object entering
Eq.~(\ref{int}), therefore we want to concentrate our studies on it. The
counterpart of $\rho(s)$ on the experimental side is the hadronic spectral
density $\rho^{\rm had}(s)$ which can be measured in $\tau$ decays in the
finite squared energy interval $[0,M_\tau^2]$ with
$M_\tau=1.777\GeV$~\cite{PDG}. The hadronic spectral density
$\rho^{\rm had}(s)$ is a rapidly varying function in the vicinity of
resonances (for the experimental data set in the vector channel see e.g.\
Fig.~\ref{fig1}), therefore there is no hope that one is able to compute this
function theoretically pointwise within perturbation theory. This fact is
related to the so-called confinement (or hadronization) problem. Instead of a
pointwise description of the spectrum at low energies, the appropriate
quantities to be analyzed theoretically in perturbative QCD are the moments or
integrals of the spectrum with a complete set of weight functions.

We define moments of the spectral density by
\begin{equation}\label{intmom}
M_l=(l+1)\int_0^{M_\tau^2}\rho(s)\frac{s^lds}{(M_\tau^2)^{l+1}}=:1+m_l.
\end{equation}
The factor $(l+1)$ in the definition of the moments is chosen to have all
contribution of the parton part uniformly normalized to unity, as it is
written explicitely in Eq.~(\ref{intmom}). This corresponds to the parton
contribution which is independent of QCD interactions. Equivalently one can
say that all measures
\begin{equation}\label{intmommem}
(l+1)\frac{s^l}{(M_\tau^2)^{l+1}}ds=d\pfrac{s}{M_\tau^2}^{l+1}
\end{equation}
defined on the interval $[0,M_\tau^2]$ are normalized to $1$ for the volume of
the integration space which in this case is the interval $[0,M_\tau^2]$. The
quantities $m_l$ defined in Eq.~(\ref{intmom}) are related to the interaction.
They are of real interest because they contain a wealth of information about
the hadronic spectral density $\rho^{\rm had}(s)$ and are sufficient for
theoretical studies.

Theoretical calculations of the moments can be done within OPE which contain
perturbative and power corrections. The perturbation theory can then be
further improved using methods of summation. We consider these steps in turn.

\section{Finite-order perturbation theory}
In this section we start the analysis of the moments given in
Eq.~(\ref{intmom}) with the consideration of the ordinary perturbative part of
the theoretical spectrum or Adler's function $D(Q^2)$. The theoretical
prediction for the function $D(Q^2)$ has been calculated with a very high
degree of accuracy within perturbation theory (see e.g.\
Refs.~\cite{phys_report,eek2,eek2c}). 
 
In the massless limit for light quarks ($u,d$) the perturbation theory
expression for Adler's function is the same for both the axial and the vector
channel. In the $\overline{\rm MS}$-scheme it reads
\begin{equation}\label{adler}
D(Q^2)=1+\frac{\alpha_s}\pi+k_1\pfrac{\alpha_s}\pi^2
  +k_2\pfrac{\alpha_s}\pi^3+k_3\pfrac{\alpha_s}\pi^4+O(\alpha_s^5)
\end{equation}
where the running coupling $\alpha_s$ is normalized at the scale $\mu^2=Q^2$,
$\alpha_s\equiv\alpha_s(Q^2)$. Here
\begin{eqnarray}\label{numbers}
k_1&=&\frac{299}{24}-9\zeta(3)\approx 1.63982,\nonumber\\
k_2&=&\frac{58057}{288}-\frac{779}4\zeta(3)+\frac{75}2\zeta(5)\approx 6.37101
\end{eqnarray}
for $n_f=3$ light flavours. $\zeta(z)$ is Riemann's $\zeta$ function. The
fourth-order $\overline{\rm MS}$-scheme coefficient $k_3$ in Eq.~(\ref{adler})
is related to the divergent parts of five-loop diagrams and is not yet known
at present. Eqs.~(\ref{adler}) and~(\ref{numbers}) constitute the set of
theoretical information necessary for a perturbation theory analysis of the
$\tau$ system. The analysis of finite-order perturbation theory was presented
in Ref.~\cite{Korner:2000ef}.

Starting with Adler's function given in Eq.~(\ref{adler}) we deduce the
spectral density $\rho(s)$ within the $\overline{\rm MS}$-scheme,
\begin{equation}\label{rhoms}
\rho(s)=1+\frac{\alpha_s(s)}\pi+k_1\pfrac{\alpha_s(s)}\pi^2
  +\left(k_2-\frac{\pi^2}3\beta_0^2\right)\pfrac{\alpha_s(s)}\pi^3+\ \ldots
\end{equation}
where the term proportional to $\pi^2$ is a result of the analytic
continuation from the Euclidean domain, $\beta_0$ is the leading coefficient
of the $\beta$-function.

The first step in studying the perturbative part of the moments $m_l$ in
Eq.~(\ref{intmom}) is to fix our criteria on how to decide which weight
functions are theoretically preferable. In ordinary (or finite order)
perturbation theory a natural criterion is the explicit convergence of the
perturbation series, i.e.\ the pattern of numerical behaviour of consecutive
terms of the perturbation series. In general, one expects the convergence of
the perturbation series to be better for moments which are more
`perturbative', i.e.\ sensitive to larger scales. However, the explicit
convergence can be concealed by the use of a particular scheme, for instance
by defining the coupling constant in the $\overline{\rm MS}$-scheme which is
most widely used. Therefore, the criterion of explicit convergence is scheme
dependent. Besides physical reasons of the applicability of perturbation
theory to a given observable the explicit convergence for the moments
(numerical structure of the perturbation theory series with finite number of
terms) can look better or worse depending on the choice of the expansion
parameter in a given scheme. This is an artefact of the definition of the
coupling (renormalization or subtraction).

To eliminate such an artificial influence of the scheme definition on physical
conclusions one should use scheme independent criteria of theoretical
calculability within perturbation theory. The invariant content of the
investigation of the spectrum in perturbative QCD, independent of any
definition of the charge or the coupling parameter, is the simultaneous
analysis of a set of moments. Technically, an efficient way to do this is to
factor out all unrelated constants by redefining the charge and introducing an
effective coupling which absorbs all constant terms of the perturbative
expansion for the polarization function or the spectral density. The purpose
of the effective coupling is to get rid of artificial scheme-dependent
constants in the perturbation theory expressions. Therefore, in order to
analyze a set of moments $m_l$ in the most transparent way (see e.g.\
Refs.~\cite{prl,brodsky1,brodsky}) we express the spectral density in terms of
an effective coupling for the $\tau$ system.

Note that the use of an effective coupling is a technical trick which
simplifies the analysis. Such a parameter is necessary in order to be able to
assess the explicit convergence of the perturbation series. The really
invariant measure is given by the mutual relations between moments while the
effective coupling is still a parameter of the theory to which one cannot
prescribe any special physical meaning. Despite this fact, the introduction of
a natural internal coupling parameter allows one to extend the perturbation
theory series available for the description of the relations between
observables by one more term as compared to the analysis in e.g.\ the
$\overline{\rm MS}$-scheme (see e.g.\ Refs.~\cite{prl,renRS}). 

In this section we work within finite-order perturbation theory (FOPT) and
define an effective coupling $a_M(s)$ on the physical cut for sufficiently
large values of $s$ by the relation
\begin{equation}\label{defofa}
a_M(s)=\frac{\alpha_s(s)}{\pi}+k_1\pfrac{\alpha_s(s)}\pi^2
  +\left(k_2-\frac{\pi^2}{3}\beta_0^2\right)\pfrac{\alpha_s(s)}{\pi}^3+\ \ldots
\end{equation}
such that
\begin{equation}\label{rhobyeff}
\rho(s)=1+a_M(s).
\end{equation}
The subscript ``$M$'' stands for a Minkowskian definition of the effective
coupling, i.e.\ the definition on the physical cut. The decomposition of the
spectral density in Eq.~(\ref{rhobyeff}) reflects the fact that within
perturbation theory the correlator contains the parton part which is
independent of $\alpha_s$. All the constants that may appear in the
perturbation theory expression for the spectral density $\rho(s)$ due to a
particular choice of the renormalization scheme are absorbed into the
definition of the effective charge (see e.g.\
Refs.~\cite{effsch,ksch,kksch,effDh}), so that only effects of the running of
the coupling itself are left. The solution of the evolution equation for the
effective coupling,
\begin{equation}\label{rengroup}
s\frac{da_M(s)}{ds}=\beta(a_M(s))=-a_M(s)
  \left(\beta_0a_M(s)+\beta_1a_M(s)^2+\beta_2a_M(s)^3+O(a_M(s)^4)\right)
\end{equation}
resulting from the renormalization group analysis of the correlator with a
given effective $\beta$-function $\beta(a)$ can be obtained by quadrature,
\begin{eqnarray}\label{run}
a_M(s)&=&a_M+\int_{M_\tau^2}^s\beta(a_M(s'))\frac{ds'}{s'}\nonumber\\
  &=&a_M+\beta_0La_M^2+(\beta_1L+\beta_0^2L^2)a_M^3
  +(\beta_2L+\frac52\beta_1\beta_0L^2+\beta_0^3L^3)a_M^4+O(a_M^5)
\end{eqnarray}
where $a_M=a_M(M_\tau^2)$ and $L=\ln(M_\tau^2/s)$. Moments of the spectral
density with different weight functions emphasize different regions of the
evolution trajectory of the renormalization group equation. Because we are
interested in the running of the effective coupling, we can also use this
solution in terms of an expansion in $L$ which is more relevant for our
particular setup, instead of an expansion in $a_M$ (which is standard).
At any given order of perturbation theory for the effective or for the exact
function $\beta(\alpha)$ the running coupling $a_M(s)$ is given by the
evolution translation of the initial value $a_M$ according to the
renormalization group equation. The result reads
\begin{eqnarray}\label{taylor}
a_M(s)&=&a_M-\beta(a_M)L+\frac12\beta(a_M)\frac{\partial\beta(a)}{\partial a}
  \Bigg|_{\rlap{$\scriptstyle a=a_M$}}L^2
  -\frac16\beta(a_M)\frac{\partial}{\partial a}
  \beta(a_M)\frac{\partial\beta(a)}{\partial a}
  \Bigg|_{\rlap{$\scriptstyle a=a_M$}}L^3+O(L^4)\nonumber\\
 &=&\exp\left(-L\beta(a)\frac{\partial}{\partial a}\right)a\,\Bigg|_{a=a_M}
\end{eqnarray}
where the solution of Eq.~(\ref{rengroup}) is written in a symbolic operator
form. Note that this expression is useful for the so-called 't~Hooft scheme
where, by definition, the $\beta$-function of perturbation theory is given by
the concise expression
$\beta(\alpha)=-\beta_0(\alpha/\pi)^2-\beta_1(\alpha/\pi)^3$~\cite{hoosch}.

Defining the effective coupling $a_M(s)$ directly through the spectrum
$\rho(s)$ itself one obtains perturbative corrections to the moments in
Eq.~(\ref{intmom}) only because of running. Without running (as, for instance,
in the conformal limit of QCD with a vanishing $\beta$-function or at the
infrared fixed point) one would have  
\begin{equation}\label{norunmom}
M_l=1+a_M(M_\tau^2)\quad\mbox{or}\quad m_l=a_M(M_\tau^2)
\end{equation}
with $m_l$ defined in Eq.~(\ref{intmom}). In this situation the whole
theoretical description of the physics of the $\tau$ system in the massless
approximation (i.e.\ without strange particles and including only perturbative
corrections without possible power corrections) would reduce to the
determination of a single number $a_M(M_\tau^2)$ and consequently there would
be no problems with the convergence of the perturbation theory series. This
reminds one of the situation in QED where the coupling is defined through the
subtraction on the mass shell and simply gives the cross section in the
Thomson limit. But because of the running of $a_M(s)$ (which is important
numerically because both $a_M(M_\tau^2)$ and $\beta_0$ are large), different
observables, i.e.\ different moments of the spectral density, generate
different perturbation series from the original object $\rho(s)$ in
Eq.~(\ref{rhobyeff}) and this difference is now within the reach of
experiments. Thus, moments just allow one to study the evolution or
$\beta$-function of the effective coupling $a_M(s)$. We remind the reader that
in this section we discuss only the perturbative part of the theoretical
spectrum where power corrections are neglected.
 
The contributions of powers of logarithms (from Eqs.~(\ref{run})
or~(\ref{taylor})) to the normalized moments in Eq.~(\ref{intmom}) are given
by
\begin{equation}\label{logs1}
(l+1)\int_0^{M_\tau^2}\ln\pfrac{M_\tau^2}s\frac{s^lds}{(M_\tau^2)^{l+1}}
  =\frac1{l+1}
\end{equation}
and
\begin{equation}\label{logs2}
(l+1)\int_0^{M_\tau^2}\ln^2\pfrac{M_\tau^2}s\frac{s^lds}{(M_\tau^2)^{l+1}}
  =\frac2{(l+1)^2}.
\end{equation}
A general formula for an arbitrary (integer) power of the logarithm reads
\begin{equation}\label{logs}
(l+1)\int_0^{M_\tau^2}\ln^p\pfrac{M_\tau^2}s\frac{s^lds}{(M_\tau^2)^{l+1}}
  =\frac{p!}{(l+1)^p}.
\end{equation}
Therefore, at any fixed order of the perturbation series expansion the effects
of running die out for large values of $l$ improving the (explicit) asymptotic
structure of the perturbative series for the moments in Eq.~(\ref{intmom}).
This is obvious and expected in QCD with its property of asymptotic freedom
because the moments with weight functions $s^l$ suppress the infrared (small
$s$) region of integration where perturbation theory is not applicable. Note
that for a given order of perturbation theory the real effects of running are
connected with powers of logarithms and not with powers of the coupling $a_M$.
Therefore one computes an expansion of the spectral density $\rho(s)$ as a
Taylor series with the initial value at $M_\tau$,
\begin{equation}
\rho(s)=\rho_0(M_\tau)+\rho_1(M_\tau)L+\rho_2(M_\tau)L^2+\ldots
\end{equation}
It is just the coefficients of this Taylor series (coefficients
$\rho_i(M_\tau)$ of the powers of logarithms) that are calculated by a
perturbative expansion in the coupling constant. In the effective scheme
using $a_M$ one finds 
\begin{eqnarray}
\rho_0(M_\tau)&=&a_M,\nonumber\\
\rho_1(M_\tau)&=&-\beta(a_M),\nonumber\\
\rho_2(M_\tau)&=&\frac12\beta(a_M)\frac{\partial\beta(a_M)}{\partial a_M},\
  \ldots
\end{eqnarray}
while for the $\overline{\rm MS}$-scheme one would have
\begin{equation}
\rho_0(M_\tau)=\frac{\alpha_s(M_\tau)}\pi+k_1\pfrac{\alpha_s(M_\tau)}\pi^2
  +\left(k_2-\frac{\pi^2}3\beta_0^2\right)\pfrac{\alpha_s(M_\tau}\pi^3+\ \ldots
\end{equation}
The technical advantage of introducing the effective scheme becomes obvious.
The effects of running as given by powers of logarithms can be obtained in a
concise form as explicit functions of $a_M$ from Eq.~(\ref{logs}) if the
$\beta$-function is known. Note that only for the leading order
$\beta$-function the powers of logarithms coincide with the powers of the
coupling $a$. 

The behaviour at fixed $l$ and large $p$, i.e.\ the convergence (or asymptotic
structure) of the perturbation series with an infinite number of terms for the
moments is another story. This question was intensively discussed in the
literature. We present our analysis of this problem in
Sec.~\ref{sec:integofrunn}.

Therefore, the set of moments in Eq.~(\ref{intmom}) has a simple and clear
structure of perturbative convergence in finite-order perturbation theory:
larger values of $l$ are better behaved from the perturbation theory point of
view. However, in practice and for an efficient comparison with experimental
data the choice of weight functions is also dictated by the precision of the
available experimental data. With this constraint taken into account,
$s^l$-moments for large values of $l$ are not welcome from an experimental
point of view. They are dominated by the contributions coming from the
high-energy part of the $\tau$ decay spectrum (and therefore show a better
perturbative convergence) while the experimental accuracy for the moments
basically deteriorates with increasing $l$ because poorly known contributions
close to the high-energy end of the interval are enhanced. Therefore, with
respect to the precision aimed for the the structure of the set of the
$s^l$-moments on the experimental side is opposite to that on the theory side:
for larger $l$ the precision for the calculation of moments using experimental
data is worsening fast.

To balance the precision requirements from the experimental and theoretical
sides one can consider a modification of the set of moments (see e.g.\
Refs.~\cite{malt,schil}). Note that in general any polynomial in $s$ can be
used as a weight function for a finite interval. In a sense a complete system
of orthogonal polynomials (like the Jakobian polynomials $P^{(a,b)}(x)$ with
the rather general weight function $x^a(1-x)^b$ which are orthogonal on the
finite interval $(0,1)$~\cite{bookY}) could provide an exact expansion of the
function $\rho(s)$ in term of its moments and allows for its full
reconstruction of the special shape of the coefficients of the expansion
(supposing that $\rho$(s) is a continuous function). Therefore, the particular
choice of weight functions is determined by the efficiency of solving concrete
problems. Actually, the basis $\{s^l;l=0,\ldots,\infty\}$ is complete for
continuous functions on the interval $[0,M_\tau^2]$, and the choice of this
basis allows for the most direct evaluation of the structure of the moments
from the perturbation theory point of view which is the most relevant for the
theoretical analysis. At $s=M_\tau^2$ the spectrum is quite perturbative and
for large values of $l$ the $s^l$-moments are basically determined by
$\rho(M_\tau^2)$.

\begin{figure}
\begin{center}
\epsfig{figure=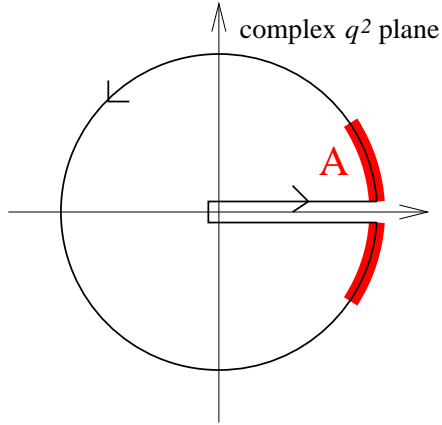, scale=0.7}
\caption{\label{fig2}contour of integration in the complex $q^2$-plane}
\end{center}
\end{figure}

As the simplest modification done in order to suppress experimental errors
from the high-energy end of the spectrum, we introduce the system of modified
moments with weight functions
\begin{equation}\label{modweight}
w_{kl}(s)=\frac{(k+l+1)!}{k!l!}
  \left(1-\frac{s}{M_\tau^2}\right)^k\pfrac{s}{M_\tau^2}^l.
\end{equation}
Even in the case $l=0$ we can consider the modified moments as mixed moments,
because they are just linear combinations of the direct moments with weight
functions $s^l$ as given in Eq.~(\ref{intmom}). The modified moments
\begin{equation}\label{intmomkl}
M_{kl}=\frac{(k+l+1)!}{k!l!}\int_0^{M_\tau^2}
  \left(1-\frac{s}{M_\tau^2}\right)^k\pfrac{s}{M_\tau^2}^l
  \frac{\rho(s)ds}{M_\tau^2}\equiv 1+m_{kl}
\end{equation}
are normalized to unity. Within the set given in Eq.~(\ref{intmomkl}) the best
choice from the experimental point of view is to use large values for $k$ and
small values for $l$. This choice was also advocated to be justified
theoretically as improving the precision based on
the integration over the contour in the complex $q^2$-plane (see
Fig.~\ref{fig2}). The reasoning was that the weight functions
$(1-s/M_\tau^2)^k$ suppress the contribution of that part of the contour that
is close to the real positive semi-axis where OPE is not applicable (region A
in Fig.~\ref{fig2}) which in turn, according to standard wisdom, can improve
the accuracy of theoretical predictions. Below we give arguments that this is
not the case: the weight functions $(1-s/M_\tau^2)^k$ simply ruin the
perturbative structure of the moments for large values of $k$. Let us
demonstrate this in more detail.

Within finite-order perturbation theory the integration over the
contour~\cite{cont,cont1,cont2} is completely equivalent to the integration
over the cut along the positive semi-axis because of the analytic properties
of the functions $\ln^p(-M_\tau^2/q^2)$ occuring in the expansion of the 
correlator. This can easily be seen from the spectral density according to
Eq.~(\ref{rhobyeff}). All polynomial moments can be rewritten as contour
integrals in the complex $q^2$-plane. Therefore, for finite-order perturbation
theory it makes no difference which particular representation for the moments
is used: integrals over the contour in the complex $q^2$-plane and integrals
over the cut along the positive semi-axis are mathematically equivalent at any
order of perturbation theory. Power corrections and infinite summation in
perturbation theory will be discussed later on.

\begin{figure}
\epsfig{figure=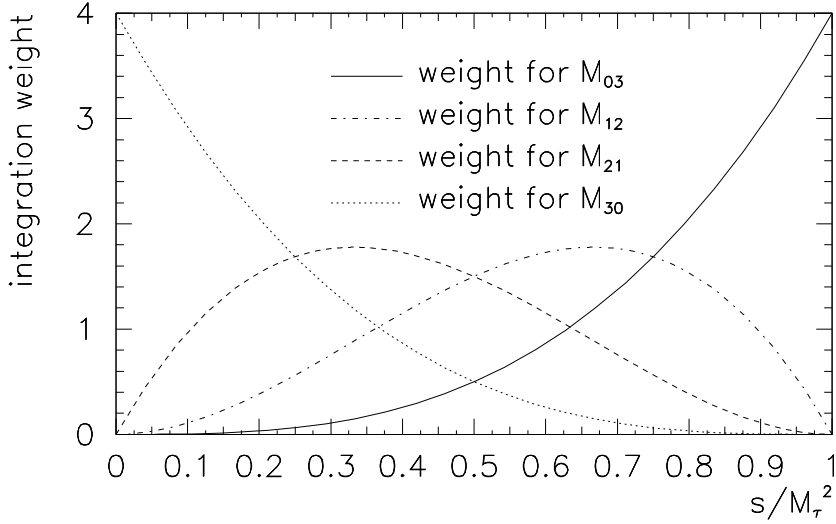, scale=0.7}
\caption{\label{fig3}Different weight functions $w_{kl}(s)$ for $(k,l)=(3,0)$,
  $(2,1)$, $(1,2)$, and $(0,3)$.}
\end{figure}

In the framework of finite-order perturbation theory we will discuss in the
following our statement that weight functions $(1-s/M_\tau^2)^k$ with large
$k$ are a bad choice for moments from the perturbation theory point of view,
i.e.\ they are strongly nonperturbative. Indeed, the weight function in
Eq.~(\ref{modweight}) has its maximum value at $s_{\rm max}=M_\tau^2l/(l+k)$
(see Fig.~\ref{fig3}). The integral in Eq.~(\ref{intmomkl}) is dominated by
contributions from around this value $s_{\rm max}$ for a smooth function
$\rho(s)$ as it is given by perturbation theory. The disadvantage of choosing
such moments is that the factor $(1-s/M_\tau^2)^k$ strongly enhances the
infrared region of integration where perturbation theory is not valid. This
can be seen by looking at the explicit convergence of the perturbation series
for the moments~\cite{Korner:2000wd}.

To show this explicitly, we consider the integration of the running coupling
with the modified weight functions in Eq.~(\ref{modweight}). The analogues of
Eqs.~(\ref{logs1}), (\ref{logs2}), and~(\ref{logs}) are necessary in order to
find the behaviour of contributions due to logarithms (see
Eq.~(\ref{taylor})). The values of the coefficients of such contributions can
be readily found in a concise form for arbitrary values of $k$ at any giving
finite order of perturbation theory (for a given power of the logarithm). For
instance, the contribution of the term proportional to $L=\ln(M_\tau^2/s)$ is
given by
\begin{equation}\label{logalt}
(k+1)\int_0^{M_\tau^2}\left(1-\frac{s}{M_\tau^2}\right)^k
  \ln\pfrac{M_\tau^2}s\frac{ds}{M_\tau^2}=\sum_{j=1}^{k+1}\frac1j
  =\gamma_E+\psi(k+2)
\end{equation}
where $\gamma_E$ is Euler's constant and $\psi(z)$ is the digamma function.
In contrast to Eq.~(\ref{logs1}) the contributions in Eq.~(\ref{logalt})
increase as $\ln(k)$ for large $k$. so the coefficients of the perturbation
series for the moments grow relatively larger with increasing $k$. The
contribution of the $L^2$-term reads
\begin{equation}\label{logalt2}
(k+1)\int_0^{M_\tau^2}\left(1-\frac{s}{M_\tau^2}\right)^k
  \ln^2\pfrac{M_\tau^2}s\frac{ds}{M_\tau^2}
  =\left(\sum_{j=1}^{k+1}\frac1j\right)^2
  +\sum_{j=1}^{k+1}\frac1{j^2}=(\gamma_E+\psi(k+2))^2+\psi'(k+2)
\end{equation}
where $\psi'(z)$ is the first derivative of the digamma function. The
coefficients in Eq.~(\ref{logalt2}) grow as $\ln^2(k)$ for large $k$ which has
to be compared with Eq.~(\ref{logs2}). For higher powers of $L$ the
expressions become too lengthy to be presented here. The asymptotic behaviour
of the perturbation theory coefficients for large $k$ and fixed $p$ is totally
different from the behaviour obtained for the $s^l$-moments, the direct
moments from Eq.~(\ref{intmom}). 

The property of convergence of the resulting perturbation series for the
moments can be reformulated in the language of effective scales for the
moments themselves. This is similar to the approach where moments are
reexpressed by moments~\cite{prl}. Indeed, let us take the expansion of
Eq.~(\ref{run}) up to next-to-leading order,
\begin{equation}\label{run10}
a_M(s)=a_M+\beta_0La_M^2+O(a_M^3)=a_M+\beta_0a_M^2\ln(M_\tau^2/s)+O(a_M^3)
\end{equation}
Then for the $s^l$-moments (i.e.\ the moments $M_{0l}$ of Eq.~(\ref{intmomkl}))
one has
\begin{equation}
m_{0l}=a_M+\frac{\beta_0}{l+1}a_M^2+O(a_M^3)
\end{equation}
which translates into the effective scale $M_\tau^2e^{-1/(l+1)}$ through the
relation
\begin{equation}
a_M+\frac{\beta_0a_M^2}{l+1}+O(a_M^3)=a_M(M_\tau^2e^{-1/(l+1)})+O(a_M^3).
\end{equation}
On the other hand, for the $(M_\tau^2-s)^k$-moments given by
Eq.~(\ref{intmomkl}) with large values for $k$ and $l=0$ the result of
integrating Eq.~(\ref{run10}) reads 
\begin{equation}
m_{k0}=a_M+\beta_0a_M^2\ln(k)+O(a_M^3)=a_M(M_\tau^2/k)+O(a_M^3).
\end{equation}
This leads to the effective scale $M_\tau^2/k$ for large $k$. The results of
these explicit calculations agree with a qualitative estimate based on the
observation that the essential region of integration where integrals for the
moments are saturated for reasonably smooth functions $\rho(s)$
(which are just powers of logarithms in finite-order perturbation theory) is
located around $s_{\rm max}=M_\tau^2l/(l+k)$. Obviously, the quantity
$a_M(M_\tau^2/k)$ cannot be evaluated in perturbative QCD for large values of
$k$. To be specific, for $k=2$ we have
\begin{equation}\label{effscale2}
m_{20}=a_M+\beta_0a_M^2\frac{11}6+O(a_M^3)=a_M(M_\tau^2e^{-11/6})+O(a_M^3)
  =a_M(0.16M_\tau^2)+O(a_M^3),
\end{equation}
and for $k=3$
\begin{equation}\label{effscale3}
m_{30}=a_M+\beta_0a_M^2\frac{25}{12}+O(a_M^3)
  =a_M(M_\tau^2e^{-25/12})+O(a_M^3)=a_M(0.12M_\tau^2)+O(a_M^3).
\end{equation}
Thus we find that ,in the language of effective scales, even a value of $k=3$
requires the calculation of the effective coupling at a low scale which is
definitely located in a nonperturbative regime. In this context we remind the
reader that the effective scale $\Lambda_M$ for the effective coupling $a_M$
as defined in Eq.~(\ref{defofa}) is given by
\begin{equation} 
\Lambda_M^2=\exp(k_1/\beta_0)\Lambda^2_{\overline{\rm MS}}
  =2.07\Lambda_{\overline{\rm MS}}^2.
\end{equation}
To fix the scale we take the value of $\Lambda_{\overline{\rm MS}}=350\MeV$ as
determined from $\tau$ decays~\cite{kr3}. Then the process dependent parameter
$\Lambda_M$ determining the effective coupling is numerically given by 
\begin{equation}
\Lambda_M^2=2.07\times(0.35\GeV)^2\approx 0.1\GeV^2.
\end{equation}
This is to be compared with the effective scale from Eqs.~(\ref{effscale2})
and~(\ref{effscale3}). For $M_\tau=1.777 \GeV$ one finds
$0.12\times M_\tau^2=0.36\GeV^2$ which is definitely located in the
nonperturbative region for the process under consideration.

This is the situation with finite-order perturbation theory which is expected
on the basis of asymptotic freedom of QCD and which is readily confirmed by a
direct calculation. As the main output of this analysis we observe the
movement of the related scale to where integrals are saturated and conclude
that $(M_\tau^2-s)^k$-moments for large $k$ cannot be calculated in
perturbation theory. Because the picture of explicit convergence due to the
effects of running can be concealed by the use of a particular scheme (the
$\overline{\rm MS}$-scheme, for instance) we have discussed the essence of the
situation in an effective scheme where only effects of running are seen for
the moments but no artificial constants related to the definition of an
expansion parameter in a specific scheme occur.

To conclude this section, we summarize our findings saying that demands from
the theory side, basically dictated by the calculability in perturbation
theory, and from the experiment side, basically dictated by the precision of
experimental data, lead to conflicting requirements for the choice of weight
functions as for where they should give the dominant contribution. Thus, one
faces the usual conflict between experimental and theoretical accuracy which
in our case is reflected by the range of $(k,l)$-values for the modified
moments that are chosen as optimal observables. Having explicit perturbation
theory formulae at hand (see e.g.\ Eq.~(\ref{adler}), (\ref{rhoms}),
(\ref{defofa}), (\ref{rhobyeff}), (\ref{rengroup}), and~(\ref{run})) one can
establish the ultimate theoretical accuracy implied by the asymptotic
character of the perturbation series for a given experimental observable with
any stated precision. This allows one to conclude which error --  experimental
or theoretical -- dominates the uncertainty of an observable related to $\tau$
decay physics. From the point of view of perturbation theory, large values of
$l$ are preferable while large values of $k$ cannot be used.

However, perturbation theory is not the end of the story. Instead, if the
convergence is not satisfactory, this is an indicator that something is going
wrong. There are contributions beyond perturbation theory, so if perturbation
theory convergence is slow, perturbation theory itself is not reliable and
nonperturbative terms become important. Nonperturbative terms for two-point
correlators can be analyzed within OPE which can (and people believe, does)
describe the spectrum at low energies better than ordinary perturbation theory.

\section{Power corrections}\label{sec:powercorr}
In this section we consider the contribution of power corrections within OPE
to the systems of moments given in Eqs.~(\ref{intmom}) and~(\ref{intmomkl}),
and the interplay between the magnitude of this contribution and the structure
of the perturbation theory series. In doing so we neglect, as usual, the weak
$\ln(Q^2)$ dependence of the coefficient functions of local operators within
OPE stemming from the anomalous dimension of the local operators to see the
net effect of a power-type behaviour. Though this is a rather common practice
in phenomenological applications, we do it for simplicity only; this
dependence can be readily taken into account (see e.g.\ the full analysis of a
realistic case below). The standard power corrections due to nonvanishing
vacuum expectation values of local operators within OPE are relatively small
for $\tau$ decays but can play a crucial role for particular observables
determined by special weight functions that lead to nonperturbative moments.
Indeed, we saw in the previous section that $(M_\tau^2-s)^k$-moments are not
computable in perturbation theory for large values of $k$. Here we will see
that one needs all power corrections which are known (or at least many terms)
to bring the experimental results for the $(M_\tau^2-s)^k$-moments in
consistency with the theoretical framework based on OPE, i.e.\ the expansion
in power corrections.

For the system of moments in Eq.~(\ref{intmom}) the contribution of power
corrections reduces to a single term of the form $(\Lambda^2/M_\tau^2)^l$
which decreases with $l$ for $\Lambda<M_\tau$ where $\Lambda$ is a typical
scale of the power corrections related to the nonperturbative scale of QCD.
This restriction for $\Lambda$ is a necessary condition for the applicability
of perturbation theory. In fact, for $\Lambda^2/M_\tau^2<1$ the perturbation
theory contribution dominates in the total result for the direct
$s^l$-moments, being a necessary condition for using the technique. It is not
quite clear, however, which numerical value should be chosen for $\Lambda$.
The choice $\Lambda\sim\Lambda_{\rm QCD}$ is just too vague. In practice, a
better choice for the scale could be $\Lambda\sim m_\rho$ or the mass of the
respective low-lying resonance in the corresponding channel as for example the
mass of the $a_1$ meson for the axial correlator (see the discussion below).

The perturbative contribution to the $s^l$-moments with large values of $l$
is saturated by high-energy contributions and therefore converges
perturbatively, the convergence becomes even better with increasing $l$.
Therefore, the $s^l$-moments are theoretically computable with a strict
control over the precision in perturbation theory. On the contrary, for
$l\sim 0$ and large values of $k$ the system of mixed moments in
Eq.~(\ref{intmomkl}) (which we call large $k$ moments in the following) is
saturated by low-energy contributions, i.e.\ basically by the contribution of
the ground-state resonance, the existence of which is a common feature of the
low-energy part of hadronic spectra. The large $k$ moments are therefore
completely nonperturbative. Power corrections to the large $k$ moments come
from many terms of OPE with local operators of high dimensionality. While
these contributions are certainly decisive for reproducing the experimental
spectrum, nothing definite can be said theoretically about such a sum of power
corrections in any realistic case because high-dimensional power corrections
are completely unknown numerically.  

An instructive example for the importance of power corrections to mixed
moments with large $k$ needed to reproduce the experimental spectrum of
hadrons at low energies is the comparison of moments for vector and axial
channels. In the massless limit the perturbation theory series for the
spectral density is the same both for the vector and axial channel leading to
identical expressions for the large $k$ moments in perturbation theory.
Experimentally, the large $k$ moments are saturated by the contributions of
the first resonances which are completely different for both channels: the
pion as a Goldstone boson in the axial channel and the $\rho$ meson in the
vector channel. Therefore, no method of summation of perturbation theory
series alone can bring the perturbation theory results for the mixed moments
with large $k$ in agreement with the experiment: if the method of summation is
a regular one, the result is the same for both channels because it sums the
same initial series. In this case perturbation theory for large $k$ moments is
in trouble and the power corrections have to provide the correct result.
Because the asymptotic regime of very large $k$ requires the consideration of
all power corrections, this example shows that large $k$ moments as in
Eq.~(\ref{intmomkl}) cannot be used within the perturbation theory framework
even if they are preferable from the experimental point of view. On the
contrary, the system of direct moments in Eq.~(\ref{intmom}) (also called $l$
moments) can be calculated reliably in perturbation theory. These moments are
saturated by perturbation theory which is also valid phenomenologically and
known since long ago (chiral invariance, Weinberg sum rules) while the
contribution of power corrections in this case is small and provides a
necessary fine-tuning for better accuracy.

Another way to understand what has been said up to now is to state that mixed
moments with large values for $k$ resolve the point-by-point structure of the
spectral density at the origin which is definitely nonperturbative while mixed
moments for large values for $l$ resolve the point-by-point structure of the
spectral density at large values for $s$ where perturbation theory is more
reliable and the standard asymptotic analysis is possible.

After these preliminary remarks we give a realistic example where the
magnitude of power corrections and perturbative terms can be seen
quantitatively. In the consideration below we use the vector channel for the
$e^+e^-$ annihilation into hadrons as an example which is relevant for the
calculation of the electromagnetic coupling constant at $M_Z$. This is the
classical channel for vector mesons ($\rho$, $\omega$). We give some details
to elaborate the real situation in QCD within OPE. This example serves as a
base for the next step -- to simplify this picture to a model which contains
only the gross features of the phenomenological structure of the spectrum but
is simple and efficient for the purpose of investigating higher order power
corrections within OPE.

\subsection{The vector channel contribution}
In this subsection we discuss the relative magnitude of perturbative and
power corrections for a particular quantity related to the two-point
correlator of hadronic currents of light quarks. The quantity which is chosen
is the value of the vacuum polarization function at the origin. This quantity
is infrared (IR) sensitive, a fact that allows one to see the interplay
between perturbative and power corrections.

For light quarks the perturbative part of the vector (and axial) correlator is
calculable for large values of $Q^2=-q^2$ within the Euclidean domain. For the
$\overline{\rm MS}$ renormalization scheme it is given by
\begin{eqnarray}\label{Pilight}
\lefteqn{\Pi^{\rm light}(\mu^2,Q^2)=\ln\pfrac{\mu^2}{Q^2}+\frac53
  +\frac{\alpha_s}\pi\left(\ln\pfrac{\mu^2}{Q^2}+\frac{55}{12}
  -4\zeta(3)\right)}\\&&
  +\pfrac{\alpha_s}\pi^2\left(\frac98\ln^2\pfrac{\mu^2}{Q^2}
  +\left(\frac{299}{24}-9\zeta(3)\right)\ln\pfrac{\mu^2}{Q^2}
  +\frac{34525}{864}-\frac{715}{18}\zeta(3)+\frac{25}3\zeta(5)\right).
  \nonumber
\end{eqnarray}
Eq.~(\ref{Pilight}) is written for $n_f=3$ active light quarks with the
effective coupling $\alpha_s\equiv\alpha_s^{(3)}(\mu^2)$. One is interested in
an estimate for the value $\Pi^{\rm light}(\mu^2,0)$. However,
$\Pi^{\rm light}(\mu^2,Q^2\to 0)$ cannot be obtained from Eq.~(\ref{Pilight})
because there is no scale for light quarks, one can therefore not avail of a
perturbative expression. Because singularities at small momenta are related to
IR problems, it suffices to modify only the IR structure of the correlator
$\Pi^{\rm light}(\mu^2,Q^2)$ in order to obtain an expression for all values
of $Q^2$. It is convenient to modify just the contribution of low-energy
states to the correlator by using the dispersion relation
\begin{equation}\label{dispNew}
\Pi^{\rm light}(Q^2)=\int_0^\infty{\rho^{\rm light}(s)ds\over s+Q^2}
\end{equation}
where dimensional regularization is understood to be used for
$\rho^{\rm light}(s)$. In fact, Eq.~(\ref{dispNew}) can be used for the bare
quantities $\Pi^{\rm light}(Q^2)$ and $\rho^{\rm light}(s)$. The modification
should be done locally (i.e.\ with a finite support in the squared energy
variable $s$ in Eq.~(\ref{dispNew})) without changing the perturbative
behaviour, in order not to affect any ultraviolet (UV) properties ($\mu^2$
dependence) of the correlator $\Pi^{\rm light}(\mu^2,Q^2)$. This requirement
is important for retaining the renormalization group properties of the
correlator. The low-energy modification of the perturbative expression for the
spectrum is inspired by experiment: at low energies there is a
well-pronounced bound state as a result of the strong interaction between
quarks. Therefore, we adopt a model of the IR modification where the
high-energy tail of the integral in Eq.~(\ref{dispNew}) is computed within
perturbation theory retaining the renormalization group structure of the
result while in the low-energy domain there is a contribution of a single
resonance.

For a generic light quark correlator in the massless perturbative
approximation one introduces the IR modification 
\begin{equation}\label{lightHadModMod}
\rho^{\rm light}(s)\to\rho_{\rm IRmod}^{\rm light}(s)
  =F_R\delta(s-m_R^2)+\rho^{\rm light}(s)\theta(s-s_0)
\end{equation}
where $F_R$, $m_R$, and $s_0$ are IR parameters of the spectrum. Note that
they may not correspond to the actual experimental numbers. Substituting the
IR modified spectrum in Eq.~(\ref{lightHadModMod}) into Eq.~(\ref{dispNew}),
one finds
\begin{eqnarray}\label{lightIRmod}
\lefteqn{\Pi_{\rm IRmod}^{\rm light}(\mu^2,0)=\frac{F_R}{m_R^2}
  +\ln\pfrac{\mu^2}{s_0}+\frac53+\frac{\alpha_s}\pi\left(\ln\pfrac{\mu^2}{s_0}
  +\frac{55}{12}-4\zeta(3)\right)}\\&&
  +\pfrac{\alpha_s}\pi^2\left(\frac98\ln^2\pfrac{\mu^2}{s_0}
  +\left(\frac{299}{24}-9\zeta(3)\right)\ln\pfrac{\mu^2}{s_0}
  +\frac{34525}{864}-\frac{715}{18}\zeta(3)+\frac{25}3\zeta(5)
  -\frac{3\pi^2}8\right).\nonumber
\end{eqnarray}
We identify $m_R$ with a mass of the low-lying resonance which is the only
input giving a scale to the problem. The IR modifying parameters $F_R$ and
$s_0$ are fixed from quark-hadron duality arguments. Notice the
$O(\alpha_s^2)$ difference between Eq.~(\ref{Pilight}) and
Eq.~(\ref{lightIRmod}) given by the additional term $-3\pi^2/8$. This is the
so-called ``$\pi^2$-correction'' (see e.g.\ Ref.~\cite{picorr}) which can be
rewritten through $\zeta(2)=\pi^2/6$. 

Next we use the OPE with power corrections that semi-phenomenologically encode
the information about the low-energy domain of the spectrum through the vacuum
condensates of local gauge invariant operators~\cite{svz} in order to
supplement the perturbative contribution. The OPE for the light quark
correlator reads
\begin{equation}\label{lightOPE}
\Pi_{\rm OPE}^{\rm light}(\mu^2,Q^2)=\Pi_{\rm IRmod}^{\rm light}(\mu^2,Q^2)
  +\frac{\langle{\cal O}_4\rangle}{Q^4}+\frac{\langle{\cal O}_6\rangle}{Q^6}
  +\ldots
\end{equation}
The quantities $\langle{\cal O}_{4,6}\rangle$ denote nonperturbative
contributions of dimension-four and dimension-six vacuum condensates. These
contributions are UV soft (they do not change the short distance properties of
the correlator) and are related to the IR modification of the spectrum. For
the purposes of fixing the numerical values of the parameters $F_R$ and $s_0$
which describe the IR modification of the spectrum one needs only the first
two power corrections $1/Q^2$ and $1/Q^4$. Note, however, that the coefficient
of the $1/Q^2$ correction vanishes because there are no gauge invariant
dimension-two operators in the massless limit. Computing the IR modified and
OPE supplemented correlation function, we find finite energy sum rules (FESR)
that alow one to fix the parameters $F_R$ and $s_0$~\cite{FESR1},
\begin{eqnarray}\label{lightsys}
F_R&=&s_0\left\{1+\frac{\alpha_s}\pi+\pfrac{\alpha_s}\pi^2
  \left(\beta_0\ln\pfrac{\mu^2}{s_0}+k_1+\beta_0\right)\right\}+O(\alpha_s^3),
  \nonumber\\
F_Rm_R^2&=&\frac{s_0^2}2\left\{1+\frac{\alpha_s}\pi+\frac{\alpha_s}\pi^2
  \left(\beta_0\ln\pfrac{\mu^2}{s_0}+k_1+\frac{\beta_0}{2}\right)\right\}
  -\langle{\cal O}_4\rangle+O(\alpha_s^3).
\end{eqnarray}
We treat $\langle{\cal O}_4\rangle$ as a small correction and take its
coefficient function as a constant (the total contribution is renormalization
group invariant). Eqs.~(\ref{lightsys}) fix $F_R$ and $s_0$ through $m_R^2$
and $\langle{\cal O}_4\rangle$. In using higher order terms in the OPE
expansion (for instance, $\langle{\cal O}_6\rangle/Q^6$) one could avoid
substituting $m_R^2$ from experiment because within the IR modification given
in Eq.~(\ref{lightHadModMod}) the IR scale is determined by the dimension-six
vacuum condensate $\langle{\cal O}_6\rangle$~\cite{FESR1}. We do not do this
because the primary purpose of the present analysis is to find the low-scale
normalization for the electromagnetic coupling and not to describe the
spectrum in the low-energy domain. The use of the experimental value for the
resonance mass $m_R^2$ makes the calculation more precise because the
numerical value for the $\langle{\cal O}_6\rangle$ condensate is not known
well (cf.\ Ref.~\cite{pivrhodual}).

The leading order solution for Eqs.~(\ref{lightsys}) (upon neglecting the
perturbative and nonperturbative corrections) is given by the parton model
result $s_0=2m_R^2$ and $F_R=s_0=2m_R^2$ which is rather precise. This
solution has been used for predicting masses and residues of the radial
excitations of vector mesons within the local duality approach where the
experimental spectrum is approximated by a sequence of infinitely narrow
resonances~\cite{FESR0}. The condensate of dimension-four operators for light
quarks is given by
\begin{equation}\label{o4numrr}
\langle{\cal O}_4\rangle
  =\frac{\pi^2}3\left(1+\frac{7\alpha_s}{6\pi}\right)\langle\frac{\alpha_s}\pi
  G^2\rangle+2\pi^2\left(1+\frac{\alpha_s}{3\pi}\right)(m_u+m_d)
  (\langle\bar uu\rangle+\langle\bar dd\rangle).  
\end{equation}
The order $\alpha_s$ corrections to the gluon condensate were calculated in
Ref.~\cite{chetglu}. We retain small corrections proportional to the light
quark masses and treat them according to the approximation of isotopic
symmetry for the light quark condensates,
$\langle\bar uu\rangle=\langle\bar dd\rangle$, which is rather precise for $u$
and $d$ quarks. The quark condensate part of Eq.~(\ref{o4numrr}) is given by
the partially conserved axial current (PCAC) relation for the $\pi$ meson,
\begin{equation}
(m_u+m_d)\langle\bar uu+\bar dd\rangle=-f_\pi^2m_\pi^2.
\end{equation}
Here $f_\pi=133\MeV$ is the decay constant of the charged pion and
$m_\pi=139.6\MeV$ is the mass of the charged pion. For the numerical standard
value $\langle(\alpha_s/\pi)G^2\rangle=0.012\GeV^4$ of the gluon
condensate~\cite{svz} and $\alpha_s/\pi=0.1$ one finds
\begin{equation}\label{num040}
\langle{\cal O}_4\rangle=\frac{\pi^2}3\left(1+\frac{7\alpha_s}{6\pi}\right)
  \langle\frac{\alpha_s}\pi G^2\rangle-2\pi^2\left(1+\frac{\alpha_s}{3\pi}
  \right)f_\pi^2m_\pi^2=0.037\GeV^4.
\end{equation}

The relation $s_0=2m_\rho^2$, where $m_\rho=768.5\MeV$ is a mass of the lowest
($\rho$ meson) resonance in the non-strange isotopic $I=1$ vector channel, is
rather precise numerically. The gluon condensate gives a small correction to
the basic duality relation $s_0=2m_R^2$ for light quarks. The same is true for
the $I=0$ channel where the lowest resonance is the $\omega$ meson with a mass
$m_\omega=781.94\MeV$. So we conclude that the model in
Eq.~(\ref{lightHadModMod}) is rather accurate if the scale is fixed by the
low-lying resonance.

\begin{figure}
\epsfig{figure=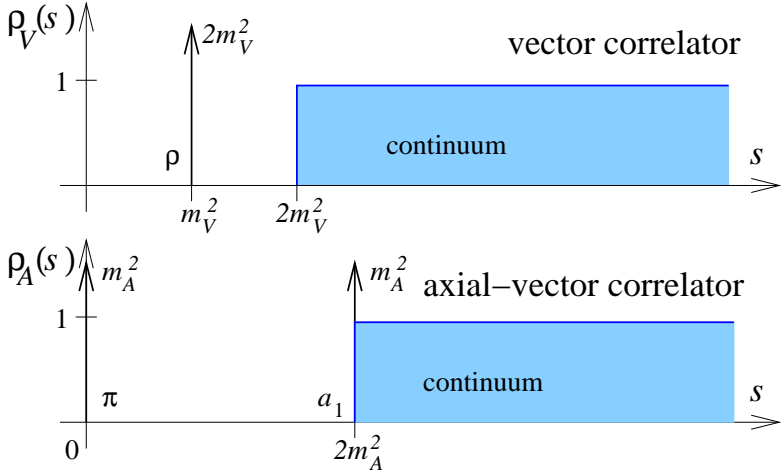, scale=0.7}
\caption{\label{fig4}Spectral densities for the models for the vector channel
  (corresponding to Eq.~(\ref{hipower}) and the axial-vector channel
  (corresponding to Eq.~(\ref{hipowerA}). The arrows indicate the narrow
  width resonances, given by distributions.}
\end{figure}

\subsection{Models for the vector and axial-vector channel}
Based on the above example we investigate models for non-perturbative effects
resulting from standard power corrections~\cite{svz} that allow us to obtain
all power corrections in a concise form. We adhere to models for the vector
and axial channels which are simple enough to be calculable and which retain
the main features of the spectra. These models are depicted in
Fig.~\ref{fig4}. With these models at hand we have all power corrections
available for the discussion of their role in moments for the spectral density
in different channels.

As a model for the vector channel we take the spectral density of the previous
subsection in a simplified form ($m\equiv m_V$),
\begin{equation}\label{hipower}
\rho_V(s)=2m^2\delta(s-m^2)+\theta(s-2m^2).
\end{equation}
By using the dispersion relation we obtain the correlation function
\begin{equation}\label{piVQ}
\Pi_V(Q^2)=\frac{2m^2}{m^2+Q^2}+\ln\pfrac{\mu^2}{2m^2+Q^2}+{\rm subtractions}
\end{equation}
with necessary subtractions. The expression in Eq.~(\ref{piVQ}) is used to
generate all power corrections. Indeed, for $Q^2\gg m^2$ one finds
\begin{equation}\label{piVQexp}
\Pi_V(Q^2)=\ln\pfrac{\mu^2}{Q^2}+\sum_{n=1}^\infty
  \left(-\frac{2m^2}{Q^2}\right)^n\left(\frac1n-\frac1{2^{n-1}}\right)
\end{equation}
where the first term is the leading order perturbative contribution in
$\alpha_s$ and the remaining terms are power corrections. Note that the
analytic properties of the expansion in Eq.~(\ref{piVQexp}) up to any finite
order are different from the exact (though model-dependent) result in
Eq.~(\ref{piVQ}). This is rather a common feature: analytic properties of 
approximations for the correlators can be different from those of the exact
result. In some instances this restricts the precision and may lead to a
misuse of approximations in areas where they do not work.

Writing the OPE for the correlator in the general form
\begin{equation}
\Pi_V(Q^2)=\ln\pfrac{\mu^2}{Q^2}+\sum_{n=1}^\infty\frac{c_n}{(Q^2)^n},
\end{equation}
for the model in Eq.~(\ref{piVQ}) we find
\begin{equation}
c_n=\left(\frac1n-\frac1{2^{n-1}}\right)(-2m^2)^n.
\end{equation}
The first two coefficients vanish, $c_1=c_2=0$. The vanishing of $c_1$ is in
full agreement with the fact that there exist no dimension-two operators in
realistic cases while the vanishing of $c_2$ means that the gluon condensate
is neglected in this model (which is justified numerically for the realistic
case of $\tau$ decays, see also the previous section). Note that these two
constraints are built-in requirements for our models. We just constructed the
models in this way. The third coefficient reads
\begin{equation}\label{third}
c_3=-\frac23m^6.
\end{equation}
Phenomenologically this coefficient is related to the vacuum expectation value
of local four-quark operators which in factorized approximation is given by
\begin{equation}\label{v6cond}
\langle{\cal O}_6^V\rangle
  =-\frac{896\pi^3}{81}\alpha_s\langle\bar qq\rangle^2.
\end{equation}
Its numerical value is approximated reasonably well by the expression in
Eq.~(\ref{third}) with $m=m_\rho$. For further comparison we give the value of
the dimension-six term in the axial channel,
\begin{equation}\label{a6cond}
\langle{\cal O}_6^A\rangle
=\frac{1408\pi^3}{81}\alpha_s\langle\bar qq\rangle^2.
\end{equation}
This vacuum expectation value is larger and has a sign opposite to that of the
vector channel.

The first few terms of the model OPE in the vector channel read explicitely
\begin{equation}
\Pi_V(Q^2)=\ln\pfrac{\mu^2}{Q^2}-\frac{2m^6}{3Q^6}+\frac{2m^8}{Q^8}
  -\frac{22m^{10}}{5Q^{10}}+\frac{26m^{12}}{3Q^{12}}+\ldots
\end{equation}
The term of dimension eight is still available in OPE as expressed through the
vacuum expectation value of local operators, even though it is very poorly
known numerically~\cite{pingro}, while higher order terms were never used in
phenomenological applications. The expression in Eq.~(\ref{piVQexp}) shows
also the actual scale of the expansion in the vector channel, $s_0=2m^2$.
The scale $\Lambda$ with $\Lambda\sim\Lambda_{\rm QCD}\sim
\Lambda_{\overline{\rm MS}}\sim 350\div 400\MeV$ or $\Lambda^2\sim 0.25m^2$
does not fit the scale of the power corrections in this model. 

There is a numerical cancellation between the resonance and continuum
contributions to the coefficients $c_n$ of the first several terms while for
higher order terms (large values of $n$) the scale $s_0=2m^2$ dominates. This
cancellation is one of the reasons for the success of the Borel sum rules for
the $\rho$ meson in the vector channel. One can go very low in the Borel
parameter $M$ and still finds power corrections to be small if only power
corrections up to dimension-six operators are included (with a nonvanishing
gluon condensate there is also an additional cancellation between the
dimension-four and dimension-six terms). Note that the expansion for Adler's
function $D(Q^2)$ (or for the renormalized correlator function $\Pi(Q^2)$)
converges for $Q^2>2m^2$. This is related to the simplicity of the analytic
structure of the correlator in the model and therefore to the simplicity of
the model spectrum. Therefore, if one is to believe that the shape of the
spectrum is really close to that of the model in Eq.~(\ref{hipower}), the
scale of power corrections should be close to the value $s_0=2m^2$. We discuss
this issue later on in more detail for a variety of different models.

Next we consider the axial part of the correlator or the axial channel.
Because of the presence of the pion, in this case the spectrum at low
energies is drastically different from the one for the vector channel. All
axial-vector resonances (with spin 1) have a finite mass. In the massless
limit there is theoretically a Goldstone mode -- corresponding to the observed
pion --  with spin zero contributing to the correlator of the axial-vector
current (this is the reason why we call it more generally axial correlator or
axial channel). The main mass scale is the mass $m_{a_1}$ of the
axial-vector meson $a_1$ which we express by $m^2_{a_1}=2m_A^2$ for further
convenience. The underlying reason for this is Weinberg's relation: at some
point one can identify the scale $m_A$ with the scale in the vector channel
which is given by the $\rho$ meson mass $m_\rho$. A possible model for the
spectrum in the axial channel reads
\begin{equation}\label{hipowerA}
\rho_A(s)=m_A^2\delta(s)+m_A^2\delta(s-2m_A^2)+\theta(s-2 m_A^2)
\end{equation}
where the first term is the pion contribution, the second one is contribution
of the $a_1$ meson, and the third represents the continuum. There is no gap
left between the second resonance and the continuum. The correlator in the
axial channel is given by
\begin{equation}\label{piAQ}
\Pi_A(Q^2)=\frac{m_A^2}{Q^2}+\frac{m_A^2}{2m_A^2+Q^2}
  +\ln\pfrac{\mu^2}{2m_A^2+Q^2},
\end{equation}
the expansion at large $Q^2$ reads
\begin{eqnarray}\label{piAQexp}
\Pi_A(Q^2)&=&\ln\pfrac{\mu^2}{Q^2}+\sum_{n=1}^\infty
  \left(-\frac{2m_A^2}{Q^2}\right)^n
  \left(\frac1n-\frac12(1+\delta_{n1})\right)\nonumber\\
  &=&\ln\pfrac{\mu^2}{Q^2}+\frac{4m_A^6}{3Q^6}-\frac{4m_A^8}{Q^8}
  +\frac{48m_A^{10}}{5Q^{10}}-\frac{64m_A^{12}}{3Q^{12}}+\ldots
\end{eqnarray}
where $\delta_{n1}$ is the Kronecker symbol. Here the contribution of the
dimension-four operator is again zero while the dimension-six contribution is
positive and larger than that in the vector channel, which is the case also in
the (model independent) OPE (see Eq.~(\ref{a6cond})). While the continuum
contribution (the logarithm and the $1/n$ part in the sum in
Eq.~(\ref{piVQexp})) remains the same, the factor $-1/2^{n-1}$ in case of the
vector channel is replaced by $-1/2$ (for $n>1$) in Eq.~(\ref{piAQexp}).
Therefore, higher order power corrections for the vector channel are dominated
by the continuum while for the axial channel they are dominated by the
resonance contributions and are generally larger -- the mass of the $a_1$
meson gives the scale both for the resonance contributions and the continuum
threshold in this particular model. However, this cannot be quantitatively
checked at present because the numerical values of the higher order
condensates are not known phenomenologically with sufficient accuracy.

It is instructive to compare the results for the model OPE expansions in the
two channels,
\begin{eqnarray}\label{aaa}
\Pi_V(Q^2)&=&\ln\pfrac{\mu^2}{Q^2}-\frac{2m_V^6}{3Q^6}+\frac{2m_V^8}{Q^8}
-\frac{22m_V^{10}}{5Q^{10}}+\frac{26m_V^{12}}{3Q^{12}}+\ldots\nonumber\\
\Pi_A(Q^2)&=&\ln\pfrac{\mu^2}{Q^2}+\frac{4m_A^6}{3Q^6}
-\frac{4m_A^8}{Q^8}
  +\frac{48m_A^{10}}{5Q^{10}}-\frac{64m_A^{12}}{3Q^{12}}+\ldots
\end{eqnarray}
The coefficients of the power corrections are different and the signs of
corresponding terms are opposite. A numerical comparison is possible if one
identifies the scales $m_A=m_V=m$ in both channels which is a rather
reasonable phenomenological approximation because of Weinberg's relation
$m_{a_1}^2=2m_\rho^2$. However, one should keep in mind that both models only
capture the gross features of the spectra while the fine details (visible at
high resolution in the energy) are different and of importance for the
numerical magnitude of higher order power corrections. Indeed, higher order
power corrections are sensitive to the fine details of the spectra and cannot
be reliably determined within rough models. Therefore, while the models
proposed in this subsection allow one to calculate power corrections up to any
order, a reasonable accuracy is expected only for low orders in $1/Q^2$ --
high-order power corrections can resolve the fine structure of the spectrum
which is not caught by the gross models. It is expected that for the first few
terms of the power expansion the accuracy is rather good while for high-order
terms it can be only an order of magnitude approximation. Still such simple
models based on the gross features of the spectrum are definitely useful for a
general analysis. Higher order terms are more sensitive to details of the
spectrum and can be predicted only with large errors. We discuss these issues
below.

The moments corresponding to the model spectral density for the vector channel
for $k=0$ read (assuming $2m^2<M_\tau^2$)
\begin{equation}\label{mom0l}
M_{0l}=1-\pfrac{2m^2}{M_\tau^2}^{l+1}\left(1-\frac{l+1}{2^l}\right).
\end{equation}
One observes that the perturbative contribution is represented by the first
term on the right hand side of Eq.~(\ref{mom0l}). The power corrections are
given in what follows. The combined perturbative and power correction
structure is a natural order for the direct $s^l$-moments. For large $l$ the
contribution of the power corrections decreases and the moments are saturated
by perturbation theory, i.e.\ if $m^2\ll M_\tau^2$, the power corrections for
the moments $M_{0l}$ die out fast.

For mixed moments $M_{kl}$ with $l=0$ and arbitrary $k$ we obtain (again for
$2m^2<M_\tau^2$)
\begin{equation}\label{mixmomk0}
M_{k0}=\left(1-\frac{2m^2}{M_\tau^2}\right)^{k+1}
  +(k+1)\frac{2m^2}{M_\tau^2}\left(1-\frac{m^2}{M_\tau^2}\right)^k.
\end{equation}
The magnitude of these moments tends to zero for large values of $k$ and
definitely cannot be represented perturbatively. Indeed, according to the
structure of OPE one should represent the moments as 
\begin{equation}\label{kdecomp}
M_{k0}=1+\Delta_{k0}
\end{equation}
where $\Delta_{k0}$ gives the contribution of the power correction terms
which are usually considered to be small. For large values of $k$, however,
one has $M_{k0}\to 0$ (as Eq.~(\ref{mixmomk0}) shows) and $\Delta_{k0}\to-1$
from Eq.~(\ref{kdecomp}), i.e.\ the moments are not given by an expansion near
the perturbation theory result, $M_{k0}=1$. The decomposition in
Eq.~(\ref{kdecomp}) is therefore useless for large values of $k$ because
$\Delta_{k0}$ is not small. For the first few orders in $k$ one has
\begin{eqnarray}
\label{exmix}
\Delta_{00}&=&\Delta_{10}\ =\ 0,\qquad
\Delta_{20}\ =\ -2\pfrac{m^2}{M_\tau^2}^3,\qquad
\Delta_{30}=-8\pfrac{m^2}{M_\tau^2}^3+8\pfrac{m^2}{M_\tau^2}^4,\nonumber\\
\Delta_{40}&=&-20\pfrac{m^2}{M_\tau^2}^3+40\pfrac{m^2}{M_\tau^2}^4
  -22\pfrac{m^2}{M_\tau^2}^5,\nonumber\\
\Delta_{50}&=&-40\pfrac{m^2}{M_\tau^2}^3+120\pfrac{m^2}{M_\tau^2}^4
  -132\pfrac{m^2}{M_\tau^2}^5+52\pfrac{m^2}{M_\tau^2}^6\nonumber\\
\Delta_{60}&=&-70\pfrac{m^2}{M_\tau^2}^3+280\pfrac{m^2}{M_\tau^2}^4
  -462\pfrac{m^2}{M_\tau^2}^5+ 364\pfrac{m^2}{M_\tau^2}^6
  -114\pfrac{m^2}{M_\tau^2}^7.
\end{eqnarray}
If $m^2\ll M_\tau^2$, the power corrections for the moments $M_{k0}$ are still
basically given by the lowest order term. If $2m^2$ is close to $M_\tau^2$ as
it is actually the case in $\tau$ decays (the conclusion is based on the
model spectrum), power corrections for the moments $M_{0l}$ are given by a
single operator of dimension $l$ and are relatively small while for the
moments $M_{k0}$ the power corrections are given by a linear combination of
all operators up to the specified order. It is instructive to rewrite the $k$
moments in terms of the actual scale $2m^2$. One has 
\begin{eqnarray}\label{exmix2}
\Delta_{00}&=&\Delta_{10}\ =\ 0,\qquad
\Delta_{20}\ =\ -\frac14\pfrac{2m^2}{M_\tau^2}^3,\qquad
\Delta_{30}\ =\ -\pfrac{2m^2}{M_\tau^2}^3
  +\frac12\pfrac{2m^2}{M_\tau^2}^4,\nonumber\\
\Delta_{40}&=&-\frac52\pfrac{2m^2}{M_\tau^2}^3
  +\frac52\pfrac{2m^2}{M_\tau^2}^4
  -\frac{11}{16}\pfrac{2m^2}{M_\tau^2}^5,\nonumber\\
\Delta_{50}&=&-5\pfrac{2m^2}{M_\tau^2}^3
  +\frac{15}2\pfrac{2m^2}{M_\tau^2}^4-\frac{33}8\pfrac{2m^2}{M_\tau^2}^5
  +\frac{13}{16}\pfrac{2m^2}{M_\tau^2}^6\nonumber\\
\Delta_{60}&=&-\frac{35}4\pfrac{2m^2}{M_\tau^2}^3
  +\frac{35}2\pfrac{2m^2}{M_\tau^2}^4
  -\frac{231}{16}\pfrac{2m^2}{M_\tau^2}^5
  +\frac{91}{16}\pfrac{2m^2}{M_\tau^2}^6
  -\frac{57}{64}\pfrac{2m^2}{M_\tau^2}^7.
\end{eqnarray}
Looking at Eq.~(\ref{exmix2}) one can conclude that keeping only the first
contribution can give a completely wrong answer for the total result for
$\Delta_{k0}$ at large $k$. Therefore, being expressed through the natural
scale, i.e.\ the scale where the analytic properties of the correlator change,
the two sets of moments behave differently. The direct moments have power
corrections in accordance with the expectation that these corrections are
powers of the relevant scale while the numerical magnitude of mixed moments is
non-predictable on general grounds without a detailed knowledge of the
spectrum at low energies. Note again that we neglect the effects of the
anomalous dimensions of local operators in our simplified consideration based
on a model spectrum which, however, is also a common practice in 
phenomenological applications.

The dominance of the resonances for the axial channel becomes obvious if we
consider the moments. For $2m_A^2<M_\tau^2$ the direct moments are given by
\begin{equation}
M_{00}=1,\qquad
M_{0l}=1+\pfrac{2m_A^2}{M_\tau^2}^{l+1}\left(1-\frac{l+1}2\right),
\end{equation}
while the mixed moments for $l=0$ read
\begin{equation}\label{mixmomk0axial}
M_{k0}=\left(1-\frac{2m_A^2}{M_\tau^2}\right)^{k+1}+(k+1)\frac{m_A^2}{M_\tau^2}
  +(k+1)\left(1-\frac{2m_A^2}{M_\tau^2}\right)^k\frac{m_A^2}{M_\tau^2}.
\end{equation}
Because of the second term in $M_{k0}$ which comes from the pion resonance
at $s=0$, the moments $M_{k0}$ will not vanish for increasing values of $k$
but will increase linearly in $k$. This is a feature worth emphasizing. For
higher and higher values of $k$ the moments $M_{k0}$ in Eq.~(\ref{mixmomk0})
go to zero if used for smooth spectral functions on the cut which is the case
if the spectral densities are calculated in perturbation theory. However, the
experimental shape of the spectrum is different, and the spectral density
partially behaves like a distribution. The total pion contribution for example
is concentrated at one point which is not affected by the decrease of the
essential support of the measure with increasing $k$. Therefore, due to the
normalization chosen the pion contribution grows with $k$. These two features
of the experimental $k$ moments, namely the decrease of the vector
contribution and the increase of the axial contribution due to the pion (as
found in Eqs.~(\ref{mixmomk0}) and~(\ref{mixmomk0axial})) are very essential
for the comparison with the theoretical description of the $\tau$-moments to
be successful. Indeed, one has
\begin{equation}\label{taumoms}
M_{kl}^\tau=N_{kl}^\tau\int_0^{M_\tau^2}\left(1-\frac{s}{M_\tau^2}
  \right)^{k+2}\pfrac{s}{M_\tau^2}^l\left(1+\frac{2s}{M_\tau^2}\right)
  \frac{\rho(s)ds}{M_\tau^2}
\end{equation}
(the normalization factor $N_{kl}^\tau$ is chosen so that $M_{kl}^\tau=1$ for
$\rho(s)=1$). Remark that the normalized $\tau$ lepton decay rate $R_\tau$ as
in Eq.~(\ref{rate}) is just $M_{00}^\tau$. The theoretical expression for the
moments changes slowly with $k$, therefore the sum of vector and axial parts
behave better than these two parts separately.

\begin{table}\begin{center}
\begin{tabular}{|l||c|c|c|c|c|c|}\hline
&$l=2$&$l=3$&$l=4$&$l=5$&$l=6$&$l=7$\\\hline
$\strut M_{0l}^{V{\rm\ ex}}$
  &$0.98692$&$0.99021$&$0.99497$&$0.99777$&$0.99909$&$0.99964$\\
$\strut M_{0l}^{V(3)}$
  &$0.98692$&$1.0000$&$1.0000$&$1.0000$&$1.0000$&$1.0000$\\\hline
$\strut M_{0l}^{A{\rm\ ex}}$
  &$1.0262$&$1.0196$&$1.0110$&$1.0055$&$1.0026$&$1.0011$\\
$\strut M_{0l}^{A(3)}$
  &$1.0262$&$1.0000$&$1.0000$&$1.0000$&$1.0000$&$1.0000$\\\hline
$\strut M_{0l}^{\rm ex}$
  &$1.0065$&$1.0049$&$1.0030$&$1.0016$&$1.0008$&$1.0004$\\
$\strut M_{0l}^{(3)}$
  &$1.0065$&$1.0000$&$1.0000$&$1.0000$&$1.0000$&$1.0000$\\\hline
\end{tabular}
\caption{\label{tab1}Moments $M_{kl}$ for $k=0$ for the vector cannel and the
  axial channel model, using the exact result $M_{0l}^{\rm ex}$ as well as the
  power series expansion up to the third order term in $1/Q^2$,
  $M_{0l}^{(3)}$. We use the estimate
  $m_A^2=m_V^2=m_\rho^2=769.3\MeV$~\cite{PDG}.}
\end{center}
\end{table}

\begin{table}\begin{center}
\begin{tabular}{|l||c|c|c|c|c|c|}\hline
&$k=2$&$k=3$&$k=4$&$k=5$&$k=6$&$k=7$\\\hline
$\strut M_{k0}^{V{\rm\ ex}}$
  &$0.98692$&$0.95745$&$0.91306$&$0.85716$&$0.79358$&$0.72591$\\
$\strut M_{k0}^{V(3)}$
  &$0.98692$&$0.94766$&$0.86915$&$0.73830$&$0.54203$&$0.26725$\\\hline
$\strut M_{k0}^{A{\rm\ ex}}$
  &$1.0262$&$1.0851$&$1.1748$&$1.2902$&$1.4256$&$1.5761$\\
$\strut M_{k0}^{A(3)}$
  &$1.0262$&$1.1047$&$1.2617$&$1.5234$&$1.9159$&$2.4655$\\\hline
$\strut M_{k0}^{\rm ex}$
  &$1.0065$&$1.0213$&$1.0439$&$1.0737$&$1.1096$&$1.1510$\\
$\strut M_{k0}^{(3)}$
  &$1.0065$&$1.0262$&$1.0654$&$1.1308$&$1.2290$&$1.3664$\\\hline
\end{tabular}
\caption{\label{tab2}Moments $M_{kl}$ for $l=0$ for the vector cannel and the
  axial channel model, using the exact result $M_{k0}^{\rm ex}$ as well as the
  power series expansion up to the third order term in $1/Q^2$,
  $M_{k0}^{(3)}$. We use the estimate
  $m_A^2=m_V^2=m_\rho^2=769.3\MeV$~\cite{PDG}.}
\end{center}
\end{table}

\begin{table}\begin{center}
\begin{tabular}{|l|rrrrrrrr|}\hline
$l=7$&$-0.00$&$-0.01$&$-0.02$&$-0.02$&$-0.03$&$-0.03$&$-0.02$&$-0.00$\\
$l=6$&$-0.01$&$-0.01$&$-0.02$&$-0.02$&$-0.02$&$-0.01$&$+0.00$&$+0.02$\\
$l=5$&$-0.01$&$-0.02$&$-0.02$&$-0.02$&$-0.01$&$+0.00$&$+0.01$&$+0.02$\\
$l=4$&$-0.01$&$-0.02$&$-0.02$&$-0.01$&$-0.00$&$+0.00$&$+0.01$&$+0.02$\\
$l=3$&$-0.01$&$-0.01$&$-0.01$&$-0.01$&$-0.00$&$+0.01$&$+0.04$&$+0.10$\\
$l=2$&$+0.02$&$+0.06$&$+0.14$&$+0.32$&$+0.66$&$+1.27$&$+2.33$&$+4.11$\\
$l=1$&$+0.02$&$+0.02$&$-0.02$&$-0.21$&$-0.71$&$-1.88$&$-4.34$&$-9.24$\\
$l=0$&$-0.01$&$-0.01$&$-0.02$&$-0.02$&$-0.01$&$+0.01$&$+0.00$&$-0.05$\\\hline
&$k=0$&$k=1$&$k=2$&$k=3$&$k=4$&$k=5$&$k=6$&$k=7$\\\hline
\end{tabular}
\caption{\label{tab3}Relative deviation of the moments $M^{\tau(3)}_{kl}$ for
  the power series up to the third order from the moments $M^\tau_{kl}$ for
  the full result containing the vector and the axial-vector channel model.
  We use the estimate $m_A^2=m_B^2=m_\rho^2=769.3\MeV$~\cite{PDG}.}
\end{center}
\end{table}

With the explicit models in Eqs.~(\ref{hipower}) and~(\ref{hipowerA}) at hand
we can make a quantitative analysis. We assume that these models give the
exact results at any order of the power expansion. Then we consider the
moments and figure out when it is enough to keep only the three terms usually
available in phenomenology in order to have a reasonable (given) accuracy.
Table~\ref{tab1} shows the situation for the $l$ moments. One sees that the
perturbation theory contribution dominates the results. For both vector and
axial correlators the accuracy of the three-term approximation is better than
$2\%$ and improves for large values of $l$ as expected.  

Table~\ref{tab2} shows the corresponding situation for the $k$ moments. Here
one sees the dominance of nonperturbative contributions to the vector and
axial correlator results. For $k=4$ the accuracy is already about $10\%$
and deteriorates fast. For the sum of vector and axial correlators the
accuracy of the three-term approximation is better as one can see from the
last line of Table~\ref{tab2}. This proves our statement about the behaviour
of the sum of vector and axial part which can also be seen from the expansions
in Eqs.~(\ref{aaa}). It is known up to the third order term from phenomenology
where the corresponding values are given in Eqs.(\ref{v6cond})
and~(\ref{a6cond}). Note, however, that the phenomenological results are
obtained only in the factorization approximation for the vacuum expectation
values of the four-quark operators. Thus, the pion contribution is crucial in
keeping a reasonable accuracy for the sum of axial and vector correlators
for the large $k$ moments. This corresponds to the local duality picture. 

Our general arguments are becoming more transparent if we consider mixed
moments $M_{kl}$ with non-zero values of $l$. This is done in Table~\ref{tab3}
where we give ratios
\begin{equation}
\frac{M_{kl}^\tau-M_{kl}^{\tau(3)}}{M_{kl}^\tau}
\end{equation}
for the $\tau$-moments defined in Eq.~(\ref{taumoms}). Then only
$\tau$-moments for large values of $l$ and small values of $k$ show
perturbative behaviour. The bottom line in Table~\ref{tab3} is drastically
different from the rest of the table, the reason being that the massless pion
only contributes to moments with $l=0$. This makes the large $k$ moments still
reasonably precise using only third-order power corrections.

\subsection{Modifications of the model for the axial channel}
Here we discuss the choice of the spectrum in the axial channel. One can
suggest forms differing from those given in Eq.~(\ref{hipowerA}). The reason
for this freedom is that we have two resonances which are essential at low
energies -- the pion and $a_1$ meson -- that bring in four free parameters:
the two residues of the resonances, the mass of the $a_1$ meson, and the
beginning of the continuum. Therefore, two requirements for the vanishing of
the power corrections $1/Q^2$ and $1/Q^4$ are not sufficient to uniquely
fix the low-energy spectrum in a simple way (with only one scale). There can
be many different additional requirements all reasonably close to the
experimental data. Note that the $a_1$ meson shows up as a broad resonance
which provides additional freedom in parameterizing its contribution to the
spectrum. For a precision application the best choice is to fix the pion
residue and the mass of the $a_1$ meson to their experimental values while
leaving its residue and the continuum threshold to be determined from OPE.
In fact, the experimental value of $f_\pi$ and simple duality requirements
give a rather good description of the axial channel.

Another possibility is to fix the residue of the $a_1$ meson to be $2m_A^2$
as in the vector channel (see Eq.~(\ref{hipower})). Then the spectrum in the
axial-vector channel reads
\begin{equation}\label{hipowera}
\rho_A(s)=2(\sqrt2-1)m_A^2\delta(s)+2m_A^2\delta(s-2m_A^2)
  +\theta(s-2\sqrt2m_A^2)
\end{equation}
which also leads to a result in which the coefficients of the first two power
corrections vanish. The third coefficient of the large $Q^2$ expansion of the
axial correlator now reads
\begin{equation}
c_3^A=\frac83(3-2\sqrt2)=0.457528<\frac23
\end{equation}
(cf.\ Eq.~(\ref{aaa})) which is smaller than in the model given by
Eq.~(\ref{hipowerA}). It is also smaller than in the vector channel which can
lead to a (slight) contradiction with the phenomenology if the scales $m_V$
and $m_A$ are identified and factorization of the four-quark condensate is
taken as being exact. Still it is numerically acceptable from the
phenomenological point of view.

\begin{figure}
\epsfig{figure=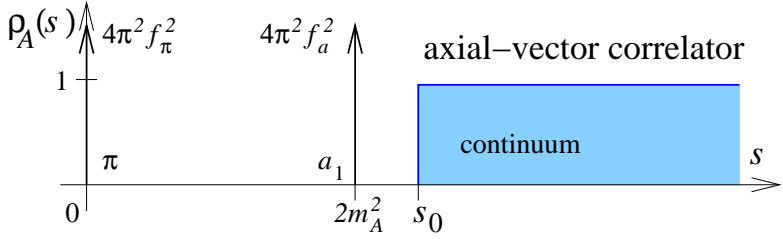, scale=0.7}
\caption{\label{fig5}Spectral density for the modified axial-vector channel
model given by Eq.~(\ref{hipoweraPar}).}
\end{figure}

Having fixed the mass of the $a_1$ meson to $2m_A^2$ one has a one-parameter
family of spectra in the axial-vector channel
\begin{equation}\label{hipoweraPar}
\rho_A(s)=4\pi^2f_\pi^2\delta(s)+4\pi^2f_a^2\delta(s-2m_A^2)+\theta(s-s_0)
\end{equation}
with $s_0$ being a free parameter. Then
\begin{equation}
4\pi^2f_a^2=\frac{s_0^2}{4m_A^2},\qquad 4\pi^2f_\pi^2=s_0-\frac{s_0^2}{4m_A^2}.
\end{equation}
The spectral density of is modified model for the axial channel is shown in
Fig.~\ref{fig5}. One constraint for the construction of the model is
$s_0>2m_A^2$. However, this constraint is quite weak. One can actually admit a
resonance as a bump in the continuum which is almost the case for the $a_1$
meson. The positivity constraint for $f_\pi^2$ is given by $s_0<4m_A^2$ and
has to be taken seriously. Note that at $s_0=4m_A^2$ the pion decouples and
does not contribute ($f_\pi=0$) to the spectrum of the correlator which is not
the case in reality. For some values of $s_0$ in the interval
$2m_A^2<s_0<4m_A^2$ the spectrum can be fixed by taking the experimental value
for $f_\pi$. A possible additional constraint is given by the value of the
dimension-six condensate. This was used for the finite-energy sum rule
analysis and turned out to be rather successful
phenomenologically~\cite{FESR1}.

The residue of the $a_1$ meson and the continuum threshold for the model in
Eq.~(\ref{hipoweraPar}) (as being expressed in terms of the pion decay
constant $f_\pi$ and the mass $m_{a_1}$ of the $a_1$ meson) are given by
\begin{equation}\label{newparaxial}
s_0=2m_A^2\left(1+\sqrt{1-4\pi^2f_\pi^2/m_A^2}\right),\qquad
  4\pi^2f_a^2=s_0-4\pi^2f_\pi^2.
\end{equation}
The consistency constraint reads
\begin{equation}
4\pi^2f_\pi^2/m_A^2\le 1.
\end{equation}
Experimentally we have $4\pi^2f_\pi^2\approx m_A^2$ which is close to the
model without a gap. Our conclusions about moments are valid for these models
as well.

The considerations obtained for the simple models lead to the conclusion that
the structure of power corrections for the moments with weight
$(1-s/M_\tau^2)^k$ is strongly infrared sensitive and nonperturbative. One
can expect that this conclusion will be retained for a realistic model (QCD),
i.e.\ that power corrections are very important and that the perturbation
theory contribution is strongly suppressed  for the moments with weight
$(1-s/M_\tau^2)^k$. This was to be expected because these moments are
saturated by resonances.

\subsection{A note on theoretical models for power corrections}
For the theoretical analysis of large $k$ moments one needs all power
corrections. These are available only in specific models. In the previous
parts of this section we have discussed models for the spectra based on the
experimentally measured shape of the spectra while taking into account only
main characteristic features of them. However, there are also theoretical
results for the spectra that are usually obtained in simplified models of
quantum field theory and which, as one believes, can have some relation to the
low-energy hadron spectra in QCD.

The old example is two-dimensional QED (QED$_2$) which was solved by
Schwinger~\cite{Schwinger}. The OPE for current correlators in QED$_2$ was
analyzed in Ref.~\cite{schwpiv}. In this model the scale of power corrections
is given by the dimensional charge $e$ (the model is super-renormalizable).
Both vector and axial-vector currents have the same scale $e^2/\pi$, in fact
they are given by a free massive field and are trivial in this respect. The
photon acquires a mass due to the anomaly. The interesting channels are that
of the scalar and pseudoscalar currents, the correlators for which are known
exactly and give an example of a really nontrivial spectrum~\cite{schwpiv}. 

There are some results concerning all-order perturbation theory diagrams
calculations in a ladder approximation~\cite{manymod}. These results can be
used for constructing realistic four-dimensional models where all power
corrections are explicitly known. The sum rule analysis of OPE in such models
can reveal some features different from the standard
phenomenlogy~\cite{pensumrule}.

\begin{figure}
\epsfig{figure=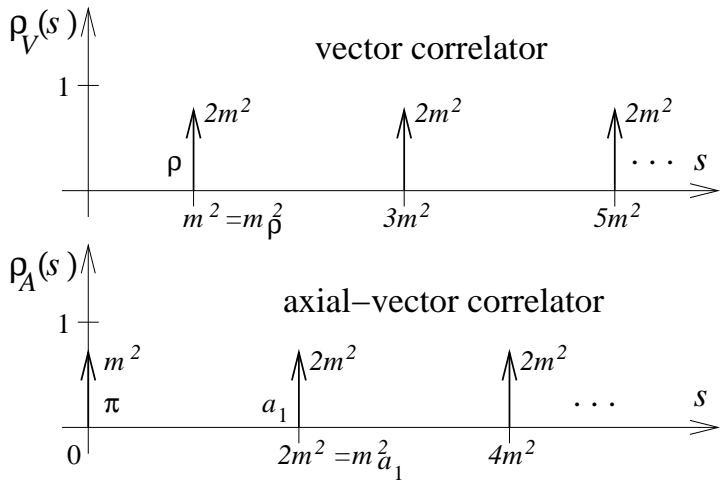, scale=0.7}
\caption{\label{fig6}Spectral density of the narrow resonance models for the
  vector and axial-vector channel as given in Eqs.~(\ref{psi})
  and~(\ref{psi2}).}
\end{figure}

A popular model is the $1/N_c$ approximation both in four and two-dimensional
Abelian and non-Abelian theories (see e.g.\ Ref.~\cite{i1nc,twoqcd}) where
the spectrum is given by infinitely narrow resonances, as shown in
Fig.~\ref{fig6}. This type of spectrum is predicted by a Regge analysis or by
local duality arguments. A recent analysis along this line was presented in
Refs.~\cite{chib,bigi}.

The model of narrow resonances inspired by the 't~Hooft model was analyzed in
QCD on the basis of local duality~\cite{FESR0}. The spectrum was studied
within the local duality approach where one has~\cite{FESR0}
\begin{equation}
m_n^2=(2n+1)m_\rho^2,\qquad f^2_n=2m_\rho^2\qquad n=0,1,\ldots
\end{equation}
for the vector channel and where $m_\rho$ is the mass of the the ground state
$\rho$ meson. For the axial-vector channel including a massless pion as
Goldstone boson (axial channel) the result of the local duality approach
coupled with the $1/N_c$ approximation reads~\cite{FESR0}
\begin{equation}\label{locdualaxial}
m_n^2=nm_{a_1}^2,\qquad f_0^2=4\pi^2f_\pi^2=\frac12m_{a_1}^2=m_\rho^2,\qquad
  f_n^2=2m_\rho^2=m_{a_1}^2,\qquad n=1,2\ldots
\end{equation}
where $m_{a_1}$ is the mass of the ground state $a_1$ meson. The structure of
the spectrum reflects the classical results on chiral symmetry and
Weinberg's relations for axial-vector and vector meson masses which is
realized if one identifies the scales in both channels,
$m_{a_1}^2=2m_\rho^2=2m^2$. This identification leads to a simplified picture
where two chains of resonances are simply shifted by an amount $m_\rho^2=m^2$.
This is the gross structure of the spectrum. Experimental results differ quite
a bit from this picture, but such details can be accounted for by using the
OPE~\cite{FESR1}.

Within the model based on the local duality approach the summation of all
resonances results in the vector correlator
\begin{eqnarray}\label{psi}
\Pi_V(Q^2)&=&\sum_{n=0}^\infty\frac{f_n^2}{m_n^2+Q^2}
  \ =\ \sum_{n=0}^\infty\frac{2m^2}{(2n+1)m^2+Q^2}\ =\nonumber\\
  &=&\sum_{n=0}^\infty\frac1{n+(Q^2+m^2)/(2m^2)}
  \ =\ -\psi\left(\frac{Q^2+m^2}{2m^2}\right)+\mbox{subtractions}.\qquad
\end{eqnarray}
Here $\psi(z)$ is the digamma function. To find the link to the OPE we
consider the large $Q^2$ behaviour of the model correlator in Eq.~(\ref{psi}).
The large $z$ expansion of the digamma function $\psi(z)$ reads
\begin{equation}
\psi(z)=\ln z-\frac1{2z}-\sum_{k=1}^{n-1}\frac{B_{2k}}{2kz^{2k}}
  +O\left(\frac1{z^{2n}}\right)
\end{equation}
where $B_{2k}$ are the Bernoulli numbers
\begin{equation}
B_{2k}=\frac{(-1)^{k-1}(2k)!}{2^{2k-1}\pi^{2k}}\zeta(2k),\qquad
B_{2k+1}=0
\end{equation}
and $\zeta(2k)$ is the Riemann $\zeta$ function. This expansion is an
asymptotic expansion. For the correlator one then obtains
\begin{equation}\label{psi1}
\Pi_V(Q^2)=\ln\pfrac{\mu^2}{Q^2}-\frac{m^4}{6Q^4}
  +0\frac{m^6}{Q^6}+\frac{7m^8}{60Q^8}+0\frac{m^{10}}{Q^{10}}+\ldots
\end{equation}
where the renormalization scale $\mu$ comes in through the subtraction term.
Note that the expression in Eq.~(\ref{psi1}) is quite different from the
result obtained in Eq.~(\ref{piVQexp}).

In the axial channel the model based on local duality (with parameters from
Eq.~(\ref{locdualaxial})) leads to the expression
\begin{eqnarray}\label{psi2}
\Pi_A(Q^2)&=&\sum_{n=0}^\infty\frac{f_n^2}{m_n^2+Q^2}
  \ =\ \frac{f_0^2}{Q^2}+\sum_{n=1}^\infty\frac{m_{a_1}^2}{m_{a_1}^2n+Q^2}
  \ =\nonumber\\
  &=&\frac{m_{a_1}^2}{2Q^2}+\sum_{n=1}^\infty\frac1{n+Q^2/m_{a_1}^2}
  \ =\ -\frac{m_{a_1}^2}{2Q^2}-\psi\left(\frac{Q^2}{m_{a_1}^2}\right)
  +\mbox{subtractions}\qquad
\end{eqnarray}
and the large $Q^2$ behaviour reads
\begin{equation}\label{psi22}
\Pi_A(Q^2)=\ln\pfrac{\mu^2}{Q^2}+\frac{m_{a_1}^4}{12Q^4}
+0\frac{m_{a_1}^6}{Q^6}
  -\frac{m_{a_1}^8}{120Q^8}+0\frac{m_{a_1}^{10}}{Q^{10}}+\ldots
\end{equation}
In this case all power corrections of dimension $2(2k+1)$ vanish because of
$B_{2k+1}=0$. This contradicts the results of OPE since there is no explicit
(symmetry) reason for such a property. At moderate orders of $n$ the structure
of the expansion of Eq.~(\ref{psi2}) as given in Eq.~(\ref{psi22}) is 
inconsistent with the asymptotic expansion expected from the OPE
(see Eqs.~(\ref{hipower}), (\ref{piVQ}), (\ref{piVQexp}), (\ref{piAQ}),
and~(\ref{piAQexp})) and, therefore, is not supported by phenomenology. Though
the model of infinitely narrow resonances is rather attractive, this
inconsistency was the reason that this model was not analyzed quantitatively
in the literature and was instead considered only as a rough approximation for
the first few moments (duality relation). Indeed, the zeroth moment is correct
(no $1/Q^2$ terms) and the first moment is correct in the approximation of a
vanishing gluon condensate contribution but the vanishing of the dimension-six
terms are definitely unacceptable phenomenologically. The high order $1/Q^2$
behaviour can be easily corrected by shifting the parameters of the first one
or two resonances. But then the simplicity of the functional form of this
model with its one-parameter dependence for the whole spectrum is lost.

The qualitative difference of the models given in Eqs.~(\ref{psi1})
and~(\ref{psi22}) from the previous case with a continuum contribution,
however, is the analytic structure of the correlators. In the narrow resonance
model one only has a single dimensional parameter $m^2$ and one would expect
the power corrections to behave as $(m^2/Q^2)^n$ with $m^2$ determining the
scale. However, the coefficients of the power corrections given by the
Bernoulli numbers grow quite fast as the large $k$ behaviour demonstrates,
\begin{equation}\label{bernasymp}
B_{2k}=4(-1)^{k-1}\sqrt{\pi k}\left(\frac{k}{\pi e}\right)^{2k}(1+o(1))
\end{equation}
(the notation $o(x^p)$ indicates contributions which fall off to zero faster
than $x^p$ for $x\to 0$). Thus, the expansion in Eqs.~(\ref{psi1})
and~(\ref{psi22}) is only asymptotic. The reason for the divergent nature of
the series of power corrections is that for the models given in
Eqs.~(\ref{psi}) and~(\ref{psi2}) there are poles located arbitrarily far away
from the origin $Q^2=0$. In this model, therefore, there is in fact an
infinite number of scales $nm^2$ for positive integers $n$. This is the reason
why perturbation theory does not work even at sufficiently large $s$. But if
one approximates a chain of resonances by a continuum starting from some
threshold $s_0$, one instead gets a finite radius of convergence of the order
$s_0$, $Q^2>s_0$ and a well working perturbation theory picture at $s>s_0$.

Let us compare the behaviour of moments for such models. Because of the
complicated argument of the $\psi$ function in Eq.~(\ref{psi}) for the vector
correlator, power corrections in the form of an expansion in $1/Q^2$ are not
directly given by the expansion of the $\psi$ function. This makes the
calculation of an arbitrary term for the vector correlator clumsy. In case of
the axial correlator this expansion is straightforward as one can see from
Eq.~(\ref{psi2}). Therefore, we shall only analyze the axial channel but will
give a functional relation that holds for both correlators in these models.
Identifying $2m^2$ with ${m_{a_1}^2}$, $m_{a_1}^2=2m^2$ one finds
\begin{equation}
\Pi_A(Q^2)=\Pi_V(Q^2-m^2)-\frac{m^2}{Q^2}
\end{equation}
or, when rewritten in the Minkowskian domain,
\begin{equation}
\Pi_A(q^2)=\Pi_V(q^2+m^2)+\frac{m^2}{q^2}.
\end{equation}
This relation is inspired by particular models for the spectra but can be
checked for experimental data as well (for Adler's functions which are
independent of subtractions).

The narrow resonance model for the vector and axial channel does not display
the difference between low and high energies. The only criterion of
perturbation theory calculability is the length of the averaging interval
while its position in terms of energies is almost unimportant. This is a
natural feature of the translation invariance of the spectra in these models.
We now discuss the moments in the axial channel in more detail. In order to
obtain simple estimates we take the averaging interval to be $5m^2$ (in ``real
life'' the averaging interval is given by $M_\tau^2=5.15m_\rho^2$ but
$5m^2=5m_\rho^2$ is a good approximation). We check the quality of the
decomposition from the perturbation theory point of view, i.e.\ we write the
result in the form
\begin{equation}\label{delkldef}
M_{kl}=1+\Delta_{kl}.
\end{equation}
The direct $l$ moments obtained for the power expansion in
Eq.~(\ref{psi22}) are (for $l>0$) given by
\begin{equation}
M_{0l}=1+(l+1)\pfrac{m_a^2}{M_\tau^2}^{2l}B_{2l}.
\end{equation}
For large values of $l$ the decomposition as given in Eq.~(\ref{delkldef}) is
useless because the coefficients are growing very fast and thus the second
term becomes large (see Eq.~(\ref{bernasymp})). Still, for odd moments the
decomposition is precise because then only one power correction is picked up
and its contribution vanishes. However, the result based on the asymptotic
value $M_{0l}=1$ is quite different from the exact answer
\begin{equation}
M_{0l}=\pfrac{m_a^2}{M_\tau^2}^{2l}+\pfrac{2m_a^2}{M_\tau^2}^{2l}.
\end{equation}
The reason for this is that the asymptotic behaviour is too crude to resolve
the position of the poles near the origin. This can be seen from the
expression for the vector channel where the appropiate expansion parameter for
power corrections is $1/(Q^2+m^2)$ instead of $1/Q^2$. In other words, the
scale $M_\tau^2$ is two small to be able to reflect any information about
remote poles which give contributions to the asymptotic behaviour at large
$Q^2$.

\begin{table}\begin{center}
\begin{tabular}{|l||c|c|c|c|c|c|c|c|}\hline
vector&$n=0$&$n=1$&$n=2$&$n=3$&$n=4$&$n=5$&$n=10$&$n=20$\\\hline
$\Delta_{0n}^V$
  &$+0.00$&$+0.00$&$-0.03$&$-0.06$&$-0.09$&$-0.13$&$-0.40$&$-0.82$\\
$\Delta_{n0}^V$
  &$+0.00$&$+0.00$&$-0.03$&$-0.06$&$-0.09$&$-0.13$&$-0.40$&$-0.82$\\
\hline\noalign{\strut\hfill}\hline
axial&$n=0$&$n=1$&$n=2$&$n=3$&$n=4$&$n=5$&$n=10$&$n=20$\\\hline
$\Delta_{0n}^A$
  &$+0.00$&$-0.04$&$-0.04$&$-0.08$&$-0.13$&$-0.19$&$-0.53$&$-0.90$\\
$\Delta_{n0}^A$
  &$+0.00$&$+0.04$&$+0.08$&$+0.16$&$+0.26$&$+0.39$&$+1.23$&$+3.20$\\\hline
\end{tabular}
\caption{\label{tab4}Deviations $\Delta_{kl}^V$ and $\Delta_{kl}^A$ (see
  Eq.~(\ref{delkldef})) of the moments for the narrow resonance vector and
  axial channel model with respect to the parton contribution
  $\Delta_{kl}^{V/A}=1$ for $k=0$ or $l=0$.}
\end{center}\end{table}

For the mixed moments in the axial channel our result are given in the lower
part of Table~\ref{tab4}. We give only numerical results because it is not
straightforward to represent the decomposition analytically. The message is
clear: both systems of moments -- direct and mixed -- are nonperturbative
which is expected for this model where for any energy region the spectrum has
nonperturbative character (close to a singularity).

An even more symmetric picture emerges in the vector channel. In this case one
does not see such a drastic difference between moments as should be expected
from the perturbation theory point of view: the situation is symmetric, the
scale of ``bumpiness'' is the same at the both ends of the spectra. This fact
makes both systems of moments almost equivalent. If we take the average
interval $6m^2$, the result for exact moments is the same for both systems of
moments because of the relation
\begin{equation}
(1-1/6)^n+(1-3/6)^n+(1-5/6)^n=(1/6)^n+(3/6)^n+(5/6)^n,
\end{equation}
as shown in the upper part of Table~\ref{tab4}. In general, for the vector
current there is just an exact symmetry of the model spectrum $s\to(2km^2-s)$
which makes the result of the calculation for both systems of moments simply
equal. Still the contribution of power corrections is different for different
systems of moments and, moreover, does not allow to restore the shape of the
exact spectrum.

This symmetry of the simplified model spectrum is definitely violated in the
realistic phenomenological spectrum, making the region of high energy
essentially different from the low energy domain. This is because of the
asymptotic freedom in QCD where at large energies the spectrum becomes a more
and more continuous function because the features of the resonances as
distributions becomes less and less pronounced. This also means that the model
for the spectrum as an infinite chain of infinitely narrow resonances does not
properly incorporate the asymptotic freedom of QCD in a sense of scale
invariance violation.

We should note here that if there is a new situation at large energies (as the
production of heavy quarks near their threshold) one should do the analysis as
if the energy were counted from the respective thresholds. It is not the
absolute value of the energy that matters but the distance from the strong
interaction region.

As a last remark we stress that even the formal knowledge of the asymptotic
series in $1/Q^2$ does not allow one to restore the spectrum point-wise. This
reflects the same situation as in phenomenological analyses, namely that the
determination of the spectral density is an improperly posed problem (inverse
problem). In our particular example one can actually observe this feature. For
the axial channel one can still figure out the spectrum from the coefficients
by restoring the full function but for the vector channel it is not
straightforward. There is actually no systematic method for doing so. Taking
just the ``leading'' asymptotics for the coefficients one can easily obtain
identical results for the axial and vector channels which is definitely an
unwelcome result as regards low-energy phenomenology. Therefore, the technique
which could produce such a result (like the Borel summation of the ``leading''
asymptotics) can be quite misleading for calculating the low-energy spectrum
as we know it from phenomenology.

\section{On the determination of the strange quark mass}\label{sec:strangemass}
In case of the $(\bar us)$ current the mass of the strange quark should be
taken into account. This was discussed recently in
Refs.~\cite{Chetyrkin:1998ej,Korner:2000wd,pichprades}.
The conclusions we have previously drawn about moments with weight function
$(1-s/M_\tau^2)^k$ for the massless case remain the same. We comment only on a
new feature of the massive correlators with the $s$ quark mass $m_s$
considered as a small correction. The occurrence of the strange quark mass
leads to an additional freedom in the choice of the scalar function for the
momentum-type analysis because the correlator is no longer transverse
(unlike in Eq.~(\ref{scalfun})). In the literature there are different
suggestions about which of the correlator parts $\Pi_q(q^2)$ and $\Pi_g(q^2)$
in Eq.~(\ref{correlator}) is better suited for the extraction of the strange
quark mass. It is worth to have a closer look at the situation. While in
present calculations radiative corrections are taken into account, the
important term is still the leading order contribution. Because $m_s$ is
small, this gives a correction which should be taken at the lowest order
available in order to have a reasonable accuracy. The leading order spectral
density for the current $\bar u\gamma_\mu(\gamma_5)s$ related to the part
$\Pi_q(q^2)$ of the correlator in Eq.~(\ref{correlator}) reads
\begin{equation}\label{qms}
\rho_q(s)=\left(1-\frac{3m_s^4}{s^2}+\frac{2m_s^6}{s^3}\right)\theta(s-m_s^2).
\end{equation}
This spectral density is perturbatively insensitive to $m_s^2$ because at
large $s$ the expression in Eq.~(\ref{qms}) starts with a term proportional to
$m_s^4$. A term of order $m_s^2$ appears only after the integration of the
spectrum over $s$ with weight functions which are hard (nonvanishing) at the
origin (at $s\sim m_s^2$ or $s=0$ for small $m_s^2$). Moments given by powers
of $s$ (as $s^l$, $l>0$) have no contribution of order $m_s^2$. Indeed, for
the zeroth order direct moment one finds
\begin{equation}\label{mq00}
M_{00}^q=\int_{m^2}^{M_\tau^2}\frac{\rho_q(s)ds}{M_\tau^2}
  =1-3\frac{m_s^2}{M_\tau^2}+3\pfrac{m_s^2}{M_\tau^2}^2
  -\pfrac{m_s^2}{M_\tau^2}^3
\end{equation}
where a term of order $m_s^2$ occurs. On the other hand, for the first moment
\begin{equation}
M_{01}^q=2\int_{m^2}^{M_\tau^2}\pfrac{s}{M_\tau^2}\frac{\rho_q(s)ds}{M_\tau^2}
  =1+3\pfrac{m_s^2}{M_\tau^2}^2-4\pfrac{m_s^2}{M_\tau^2}^3
  +6\pfrac{m_s^2}{M_\tau^2}^2\ln\pfrac{m_s^2}{M_\tau^2}
\end{equation}
there is only a logarithmically enhanced $m_s^4$ contribution. For higher
moments there are only higher order corrections, e.g.\ for $l=2$ one has
\be
M_{02}^q=3\int_{m^2}^{M_\tau^2}\pfrac{s}{M_\tau^2}^2
  \frac{\rho_q(s)ds}{M_\tau^2}
  =1-9\pfrac{m_s^2}{M_\tau^2}^2+8\pfrac{m_s^2}{M_\tau^2}^3
  -6\pfrac{m_s^2}{M_\tau^2}^3\ln\pfrac{m_s^2}{M_\tau^2}.
\end{equation}
Therefore, perturbative $s^l$-moments of $\Pi_q(q^2)$ have no sensitivity to
$m_s^2$. This is also expected for this correlator part because the mass
corrections start from power corrections and lead to no cut which is a
perturbative singularity in QCD and reliable for large $s$. Note in passing
that this is also the reason why the perturbation theory expansion for the
coefficient function of $m_s^2$ in the correlator part $\Pi_q(q^2)$ is known
by one term less than that for the part $\Pi_g(q^2)$: for the correlator part
$\Pi_g(q^2)$ only divergent parts of the four-loop diagrams are necessary
while for the part $\Pi_q(q^2)$ one has to calculate the finite contributions
of the four-loop diagrams which is beyond the scope of present computational
algorithms. 

The afore mentioned structure of the spectral density also means that there is
no way to improve the theoretical accuracy of the contribution of the mass
by making the corresponding coefficient function more perturbative: as soon as
one tries to make the coefficient function multiplying $m_s^2$ more
perturbative by introducing an additional suppression by a power of $s$ (or by
increasing $l$ for the modified moments), the contribution of the mass becomes
parametrically smaller, so instead of $m_s^2$ one has $m_s^4$ in the leading
order. When the anomalous dimension of the mass is taken into account, the
power $s$ then leads to an additional suppression by a power of $\alpha_s$.

On the other hand, the $(M_\tau^2-s)^k$-moments all have the same contribution
of order $m_s^2$ given simply by the zero order direct moment $M_{00}^q$ of
Eq.~(\ref{mq00}),
\begin{equation}
M_{k0}^q=(k+1)\int_{m^2}^{M_\tau^2}\left(1-\frac{s}{M_\tau^2}\right)^k
  \frac{\rho_q(s)ds}{M_\tau^2}
  =1-3(k+1)\frac{m_s^2}{M_\tau^2}+o\pfrac{m_s^2}{M_\tau^2}.
\end{equation}
This amplitude is used in Ref.~\cite{pichprades} where a better stability in
the $\overline{\rm MS}$ scheme was detected.

In contrast to $\rho_q(s)$, for the spectrum $\rho_g(s)$ of the amplitude
$\Pi_g(q^2)$ in Eq.~(\ref{correlator}) one has
\begin{equation}
\rho_g(s)=\left(-s+\frac{3m_s^2}2-\frac{m_s^6}{2s^2}\right)\theta(s-m_s^2)
\end{equation}
which is sensitive to $m_s^2$ for large $s$. Any perturbative moment of this
amplitude has $m_s^2$ contributions and can be used in the perturbation theory
analysis of $m_s^2$ corrections. Namely,
\begin{equation}
M_{0l}^g=(l+1)\int_{m_s^2}^{M_\tau^2}\pfrac{s}{M_\tau^2}^k\rho_g(s)
  \frac{ds}{M_\tau^2}=-M_\tau^2\left\{\frac{l+1}{l+2}
  -\frac{3m_s^2}{2M_\tau^2}+o\pfrac{m_s^2}{M_\tau^2}\right\}.
\end{equation}
Therefore, the choice of the correlator part $\Pi_g(q^2)$ is definitely more
reliable perturbatively for analyzing mass corrections, a fact that is also
discussed in Refs.~\cite{Chetyrkin:1998ej,Korner:2000wd}. However, the
experimental situation is again less favourable to the choice of $\Pi_g(q^2)$
in comparison to $\Pi_q(q^2)$. The correlator part $\Pi_q(q^2)$ contains a
contribution of the $K$ meson which is well pronounced and can be easily seen
while the amplitude $\Pi_g(q^2)$ contains only spin-one states and is poorly
measured. In order to benefit from a better experimental accuracy, the use of
the amplitude $\Pi_q(q^2)$ is favoured, but this choice leads to sum rules
which are rather similar to the pseudoscalar sum rules (see e.g.\
Ref.~\cite{chetschilms}) and does not provide new aspects for the strange
quark mass determination specific to $\tau$ decays.

\section{Integration of running effects to all orders}\label{sec:integofrunn}
In this section we discuss the technique of resumming the effects of
renormalization group caused running of the strong coupling and other
parameters (simply called ``running'' in the following) by taking the
integration path along the contour of a circle in the complex $q^2$-plane. As
was discussed earlier in Sec.~3, the integration along the contour of a circle
in the complex $q^2$-plane is completely equivalent to the integration along
the cut if finite-order perturbation theory expressions for the polarization
functions are used. This is the direct mathematical consequence of the
analytic properties of the functions $\ln^p(-M_\tau^2/q^2)$ for positive
integer values $p$. However, if the renormalization group improved
polarization function is used for the integration, some new features
appear. This was also discussed in Sec.~4. The reason for this is that the
analytic properties of resummed polarization functions are different from
those in finite-order perturbation theory.

The integration along the contour including a full renormalization group
resummation is now the most popular technique of accounting for the running of
perturbative quantities: it efficiently resums an infinite number of terms
generated by the evolution of the coupling constant~\cite{Pivtau,DP}. Clearly
it also includes some nonperturbative features when an infinite number of
terms is resummed which means that this also constitutes a special recipe for
resummation of perturbation series~\cite{pivsuppl}. The formulae for the
integration along the circle are known in a closed form as one-fold integrals
but the final integration itself is usually done numerically. An analysis of
the analytical form of these integral expressions is useful for clarifying the
real content of such a resummation technique. In doing this we discuss how our
general considerations given in the previous sections work within this
resummation technique. It should definitely give the results confirming the
physical considerations (effective scale structure) qualitatively, otherwise
the technique of resummation would be irrelevant for physical applications. As
we shall see, the resummation technique fully agrees with the conclusions
about perturbative calculability and asymptotic structure of the moments found
in the analysis on the cut in the previous sections.

\subsection{Quantities on the circle}
As in the case of the analysis on the physical cut, our main concern is to
account for the running. Therefore, the introduction of an effective coupling
is helpful. Here we introduce an effective coupling in the Euclidean domain
which requires some new notation.

We start with Adler's function in Eq.~(\ref{adler}), introducing the effective
strong coupling $\alpha_E(Q^2)$,
\begin{equation}
D(Q^2)=-Q^2\frac{d}{dQ^2}\Pi(Q^2)=1+\frac{\alpha_E(Q^2)}\pi.
\end{equation}
The effective coupling $\alpha_E(Q^2)$ obeys the renormalization group
equation (see e.g.~\cite{RG})
\begin{equation}
Q^2\frac{d}{dQ^2}\pfrac{\alpha_E(Q^2)}\pi=\beta(\alpha_E(Q^2)).
\end{equation}
For the considerations we are aiming at it is enough to use only leading order
in the $\beta$-function which contains already the bulk of the whole effect.
Effects due to higher order corrections of the $\beta$-function are really
small and do not change the basic picture. They only slightly affect the
conclusions numerically~\cite{groote}. Therefore, we consider the
renormalization group equation
\begin{equation}
Q^2\frac{d}{dQ^2}\pfrac{\alpha_E(Q^2)}\pi=-\beta_0\pfrac{\alpha_E(Q^2)}\pi^2. 
\end{equation}
The resummed correlation function reads
\begin{equation}\label{resumcorr}
\Pi(Q^2)=\ln\pfrac{\mu^2}{Q^2}+\frac1{\beta_0}\ln\pfrac{\alpha_E(Q^2)}\pi
  +{\rm subtractions}
\end{equation}
where
\begin{equation}\label{alpfunc}
\alpha_E(Q^2)=\frac{\alpha_\tau}{1+(\beta_0\alpha_\tau/\pi)\ln(Q^2/M_\tau^2)}
\end{equation}
with $\alpha_\tau=\alpha_E(M_\tau^2)$. Taking this into account and
parameterizing the contour by $Q^2=M_\tau^2e^{i\varphi}$ one obtains
\begin{equation}\label{logres}
\Pi(M_\tau^2e^{i\varphi})=-i\varphi
  -\frac1{\beta_0}\ln(1+i{\beta_0}\alpha_\tau\varphi/\pi)+\mbox{subtractions}
\end{equation}
where appropriate subtractions are added. The first term is the parton
contribution which is independent of $\alpha_\tau$. This term is readily be
taken into account. We therefore concentrate on the contribution coming from
the second term which is $\alpha_\tau$ dependent.

The analysis of this second part can now be done for the moments $M_{kl}$,
given by
\begin{eqnarray}\label{momcircBas}
M_{kl}&=&1+m_{kl}\ =\ \frac{(-1)^l}{2\pi i}\frac{(k+l+1)!}{k!l!}\oint_{|z|=1}
  \Pi(M_\tau^2z)(1+z)^kz^ldz\ =\nonumber\\
  &=&\frac{(-1)^l}{2\pi}\frac{(k+l+1)!}{k!l!}\int_{-\pi}^\pi
  \Pi(M_\tau^2e^{i\varphi})(1+e^{i\varphi})^ke^{i(l+1)\varphi}d\varphi.
\end{eqnarray}
Taking the polarization function as given in Eq.~(\ref{logres}), the first
part leads to the parton result $M_{kl}=1$ while the second,
$\alpha_\tau$-dependent part gives the moment $m_{kl}$ where
\begin{equation}\label{momcirc}
m_{kl}=\frac{(-1)^{l+1}}{2\pi\beta_0}\frac{(k+l+1)!}{k!l!}\int_{-\pi}^\pi
  (1+e^{i\varphi})^ke^{il\varphi}\ln(1+i\beta_0\alpha_\tau\varphi/\pi)
  e^{i\varphi}d\varphi.
\end{equation}
Note that the estimate of the saturation region for the moments given by
integrals like the one in Eq.~(\ref{momcirc}) is a bit trickier than in the
case of the integration on the cut. On the cut the measure is positive and is
normalized to $1$ for any type of moments with $n\ge 0$,
\begin{equation}
(n+1)\int_0^1x^ndx=1\qquad\mbox{and}\qquad(n+1)\int_0^1(1-x)^ndx=1
\end{equation}
which makes the estimate of integrals
\begin{equation}
(n+1)\int_0^1x^nf(x)dx
\end{equation}
for continuous (and rather smooth) functions $f(x)$ straightforward. In the
present case, however, the measure of the integration region is oscillating
and vanishes for constant functions ($k\ge 0$, $l\ge 0$),
\begin{equation}
\int_{-\pi}^\pi(1+e^{i\varphi})^ke^{il\varphi}
  e^{i\varphi}d\varphi=0.
\end{equation}
Therefore, some additional care in the consideration of essential regions of
integration is necessary for the formulation of moments on the contour. Note
that we have already encountered such a problem while analyzing the moments of
the narrow resonance contributions. Therefore, the intuition gained from the
situation with continuous functions generally does not work for distributions.

\subsection{Techniques for the calculation of contour moments}
Because explicit functions are given in Eqs.~(\ref{resumcorr}),
(\ref{logres}), (\ref{momcircBas}), and~(\ref{momcirc}), the discussion of the
representation and properties of the moments on the contour is a pure
mathematical problem. Since that moments on the circle (as in
Eq.~(\ref{momcirc})) are just Fourier coefficients of the correlation
functions $\Pi(Q^2)$, the technique for working with these is well-known. One
finds similarities to the evaluation of sunset diagrams for the case when the
problem is reduced to a Fourier transform.

For $\beta_0\alpha_\tau<1$ the moments in Eq.~(\ref{momcirc}) can be expanded
in a convergent series in $\alpha_\tau$. The finite radius of convergence
within this technique of resummmation is a general feature which persists in
higher orders of the $\beta$-function and was studied in some
detail~\cite{groote}. The convergence radius decreases when higher orders of
the $\beta$-function are included. If we compare the value of the effective
coupling in $\tau$ decays (not to be mistaken with the coupling within the
$\overline{\rm MS}$-scheme)
\begin{equation}
\frac{\alpha_\tau^{\rm exp}}\pi=0.14
\end{equation}
with
\begin{equation}
\frac1{\pi\beta_0}=\frac4{9\pi}=0.1415\ldots,
\end{equation}
the relation $\beta_0\alpha_\tau<1$ is still valid, though almost marginally.
However, the exact expression given in Eq.~(\ref{momcirc}) without expansion
of the logarithm provides an analytic continuation beyond the convergence
radius even when $\alpha_\tau$ lies outside the convergence radius. 

We first consider the basic moment $m_{00}$. The discussion of this basic
moment leads to the technically simplest (shortest) expressions while
containing already all features of the general case. The generalization to
other moments (and types of moments) is then straightforward.  

The most direct way of calculation is to expand the logarithm in the
integrand in Eq.~(\ref{momcirc}) in a series in $\alpha_\tau$ which gives
nothing new in comparison with the finite-order perturbation theory case.
However, one can generate all terms of the series because the function is
known. The series obtained by this procedure represents an analytic function
of the coupling constant at the origin. Therefore, for small $\alpha_\tau$
this series can be used for a numerical evaluation of the moments. This
procedure may not be the best one concerning economy of evaluation since many
terms may be needed to determine values of $\alpha_\tau$ close to the
convergence boundary. A finite piece of the convergent series gives different
accuracy and can also completely misrepresent the function if $\alpha_\tau$ is
beyond the convergence radius (which is not so dramatic for positive values of
$\alpha_\tau$, though).

One can proceed with the analysis of the moments in a different way by
constructing just an efficient computational scheme. Integrating $n$ times by
parts one obtains (we remind the reader that we use the zeroth order moment
for simplicity)
\begin{equation}\label{momexpand1}
m_{00}=\frac1{\pi\beta_0}\Bigg\{\phi+\sum_{j=1}^{n-1}(j-1)!
  \pfrac{\beta_0\alpha_\tau}{\pi r}^j\sin(j\phi)
  +\frac{(n-1)!}2\pfrac{\beta_0\alpha_\tau}\pi^n\int_{-\pi}^\pi
  \frac{e^{i\varphi}d\varphi}{(1+i\beta_0\alpha_\tau\varphi/\pi)^{n}}\Bigg\}.
\end{equation}
with the polar coordinate functions $r$ and $\phi$ (note the difference to
$\varphi$!) defined by
\begin{equation}
1\pm i\beta_0\alpha_\tau=re^{\pm i\phi},\qquad
r=\sqrt{1+\beta_0^2\alpha_\tau^2},\qquad
\phi=\arctan(\beta_0\alpha_\tau).
\end{equation}
The $n$-fold integration by parts removes a polynomial of order $n$ from the
expansion of the logarithm in Eq.~(\ref{logres}).

One can see that the result is an asymptotic expansion as the residual term,
i.e.\ the last term in Eq.~(\ref{momexpand1}) is of the formal order
$\alpha_\tau^n$. However, the obtained result is not a series expansion in the
original coupling determined in the Euclidean domain but a more complicated
system of functions related to it. The system of functions is ordered and the
asymptotic expansion is valid in the sense of Poincar\'e. The system of
functions is obtained by using the expression for the running coupling in
Eq.~(\ref{alpfunc}) in the Euclidean domain and continuing it into the complex
plane and onto the cut. When the analytic structure of the initial function is
known, asymptotic expansions which converge fast for the first few terms (as a
representation in the form of Eq.~(\ref{momexpand1})) are more useful for
practical calculations than formal convergent series that require many terms
for getting a reasonable accuracy.

The expansion in Eq.~(\ref{momexpand1}) can give a better accuracy (for some
$n$ and $\alpha_\tau$) than a direct expansion in $\alpha_s$. Indeed, this
expansion includes a partial resummation of the $\pi^2$ terms which is a
consequence of the analytic continuation~\cite{picorr}. Therefore, the
expansion can be understood as being done in terms of quantities defined on
the cut. This is evident because in some sense the procedure consists in
calculating derivatives of the spectral density on the cut. Because the region
near the real axis is important, the continuation causes a change of the
effective expansion parameter
$\alpha_\tau\to\alpha_\tau/\sqrt{1+\beta_0^2\alpha_\tau^2}$. The first term in
the expansion shown in Eq.~(\ref{momexpand1}) is just the value for the
spectral density expressed through the coupling in the Euclidean domain. It
can be considered as a change of the scheme at large values of $s$.

With the concise expression for the moments at hand one can change the form of
the residual term. The relation
\begin{eqnarray}\label{end}
\lefteqn{(n-1)!\pfrac{\beta_0\alpha_\tau}\pi^n\int_{-\pi}^\pi
  \frac{e^{i\varphi}d\varphi}{(1+i\beta_0\alpha_\tau\varphi/\pi)^n}
  \ =}\nonumber\\
  &=&2\pi e^{-\pi/\beta_0\alpha_\tau}-(n-1)!\pfrac{\beta_0\alpha_\tau}\pi^n
  \left(\int_{-\infty}^{-\pi}+\int_{\pi}^{\infty}\right)
  \frac{e^{i\varphi}d\varphi}{(1+i\beta_0\alpha_\tau\varphi/\pi)^n}
\end{eqnarray}
valid for any $n$ leads to a representation of the zeroth order moment in
the form
\begin{eqnarray}\label{momexpand2}
m_{00}&=&\frac1{\pi\beta_0}\Bigg\{\pi e^{-\pi/\beta_0\alpha_\tau}+\phi
  +\sum_{j=1}^{n-1}(j-1)!\pfrac{\beta_0\alpha_\tau}{\pi r}^j
  \sin(j\phi)\nonumber\\&&\qquad
  -\frac{(n-1)!}2\pfrac{\beta_0\alpha_\tau}\pi^n
  \left(\int_{-\infty}^{-\pi}+\int_\pi^\infty\right)
  \frac{e^{i\varphi}d\varphi}{(1+i\beta_0\alpha_\tau\varphi/\pi)^{n}}
  \Bigg\}.\qquad
\end{eqnarray}
Here an explicit nonperturbative term $e^{-\pi/\beta_0a_\tau}$ has appeared.
Eq.~(\ref{momexpand2}) and Eq.~(\ref{momexpand1}) are formally different but
actually identical. Therefore, the choice for the expansion (or
representation) for the moment is a question of calculating the residual term.
This quantitative consideration is, of course, only possible if one has a
concise expression for the function from which the series is generated as
given by Eq.~(\ref{momcirc}) in our case. Any conclusions about the precision
or the analytic structure of the sum of the series based on the terms of the
series only without a specification of the residual term are rather useless as
one can see from Eq.~(\ref{end}).

We stress that the moments in Eq.~(\ref{momcirc}) are analytic functions
of $\alpha_\tau$ for small values of the coupling $\alpha_\tau$. This means
that the non-analytic piece in Eq.~(\ref{end}) cancels the corresponding piece
in the residual term. If the residual term is dropped, the analytic structure
drastically changes depending on which representation, either
Eq.~(\ref{momexpand1}) or~(\ref{momexpand2}), is used. This demonstrates the
danger of reaching conclusions about power corrections emerging from the
extrapolation of the running to the IR region. Mathematically, or from a
theoretical point of view, a conclusion of this type is a hypothesis,
especially having in mind that such power corrections are not necessary in any
way for the consistency of the theory (in contrast to the OPE corrections
which are well-defined in perturbation theory and are motivated by, say,
explicitly distinguishing different channels). Physically, the use of such
corrections is very difficult to appreciate if higher order terms of the
expansion in $\alpha_s$ are taken into account, i.e.\ an explicit
nonperturbative term in the expression, not being special in any general way,
does not numerically differ from the higher order terms as well.

\begin{figure}
\begin{center}\epsfig{figure=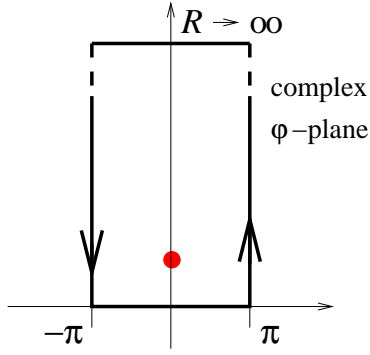,scale=0.7}
\caption{\label{fig7}the integration contour in the complex $\varphi$-plane}
\end{center}
\end{figure}

Yet another representation for the moments can be readily obtained. It is
possible to recover the form of the moments as integrals over a spectral
density (see below). But because moments are calculated as explicit functions,
we can obtain a result by using a simple change of variables in a pure 
mathematical sense. In order to obtain this new representation we go to the
complex plane in $\varphi$ (see Fig.~\ref{fig7}). As a technical trick we
first integrate once by parts to obtain
\begin{equation}
m_{00}=\frac{\alpha_\tau}{2\pi^2}\int_{-\pi}^\pi
  \frac{(1+e^{i\varphi})d\varphi}{1+i\beta_0\alpha_\tau\varphi/\pi}
\end{equation}
This representation is somewhat different from the representation in
Eq.~(\ref{momexpand1}) for $n=1$. They differ by an integral which can be
explicitly computed,
\begin{equation}
\frac{\alpha_\tau}{2\pi^2}\int_{-\pi}^\pi
  \frac{d\varphi}{1+i\beta_0\alpha_\tau\varphi/\pi}
  =\frac1{2\pi i\beta_0}\ln\pfrac{1+i\beta_0\alpha_\tau}{1-i\beta_0\alpha_\tau}
  =\frac{1}{\pi\beta_0}\arctan(\beta_0\alpha).
\end{equation}

Now we consider the integration over a rectangular contour in the complex
$\varphi$-plane. The part of the contour on the real axis from $-\pi$ to $\pi$
leads to the moments. The integral over the contour is given by the residue at
the pole $\varphi=i\pi/\beta_0\alpha_\tau$. We thus have
\begin{eqnarray}
m_{00}&=&\frac{\alpha_\tau}{2\pi^2}\int_{-\pi}^\pi
  \frac{(1+e^{i\varphi})d\varphi}{1+i\beta_0\alpha_\tau\varphi/\pi}
  \ =\nonumber\\
  &=&\frac{i\alpha_\tau}\pi\pfrac{1+e^{i\varphi}}{i\beta_0
  \alpha_\tau/\pi}\Bigg|_{\varphi=i\pi/\beta_0\alpha_\tau}\,+\nonumber\\&&
  -\frac{\alpha_\tau}{2\pi^2}\int_0^\infty\frac{(1+e^{i\pi-\xi})
  id\xi}{1+i\beta_0\alpha_\tau-\beta_0\alpha_\tau\xi/\pi}\,+
  \qquad(\varphi=\pi+i\xi)\nonumber\\&&
  +\frac{\alpha_\tau}{2\pi^2}\int_0^\infty\frac{(1+e^{-i\pi-\xi})
  id\xi}{1-i\beta_0\alpha_\tau-\beta_0\alpha_\tau\xi/\pi}\,+
  \qquad(\varphi=-\pi+i\xi)\nonumber\\&&
  +\frac{\alpha_\tau}{2\pi^2}\lim_{R\rightarrow\infty}\int_{-\pi}^\pi
  \frac{(1+e^{i\tilde\varphi-R})d\tilde\varphi}{1+i\beta_0\alpha_\tau
  \tilde\varphi/\pi-\beta_0\alpha_\tau R/\pi}.\qquad(\varphi=\tilde\varphi+iR)
\end{eqnarray}
where we have indicated the substitutions for the different parts of the path.
The last part vanishes because the integrand vanishes for $R\rightarrow\infty$
while the integration range is finite. We thus obtain
\begin{equation}
m_{00}=\frac1{\beta_0}(1+e^{-\pi/\beta_0\alpha_\tau})
  -\frac1{\beta_0}\int_0^\infty\frac{(1-e^{-\xi})d\xi}{\pi^2
  +(\xi-\pi/\beta_0\alpha_\tau)^2}.
\end{equation}
Next we replace $-\pi/\beta_0\alpha_\tau=\ln(\Lambda^2/M_\tau^2)$ and
substitute $-\xi=\ln(s/M_\tau^2)$, so that $e^{-\xi}=s/M_\tau^2$ and
$-d\xi=ds/s$. One obtains
\begin{equation}
m_{00}=\frac1{\beta_0}\left(1+\frac{\Lambda^2}{M_\tau^2}\right)
  -\frac1{\beta_0}\int_0^{M_\tau^2}\frac{(1-s/M_\tau^2)ds}{(\pi^2
  +\ln^2(s/\Lambda^2))s}.
\end{equation}
A further transformation gives
\begin{equation}\label{m00further}
m_{00}=\frac1{\beta_0}\pfrac{\Lambda^2}{M_\tau^2}
  +\frac1{\pi\beta_0}\int_0^{M_\tau^2}\arccos
  \left(\frac{\ln(s/\Lambda^2)}{\sqrt{\pi^2+\ln^2(s/\Lambda^2})}\right)
  \frac{ds}{M_\tau^2}.
\end{equation}
One easily recognizes this representation as an integration over the
singularities of $\Pi(q^2)$ in Eq.~(\ref{resumcorr}). In addition to a cut
along the positive semi-axis there appears also a part of the singularity on
the negative real $s$-axis. This part is a pure mathematical feature of the
concrete approximation chosen for $\Pi(q^2)$ and is not related to the
physical content of the problem. Indeed, if moments are written as explicit
functions we can calculate them in the way we find most convenient for a
concrete application. The result reads
\begin{equation}\label{m00result}
m_{00}=\int_{-\Lambda^2}^{M_\tau^2}\frac{\sigma(s)ds}{M_\tau^2}
\end{equation}
with
\begin{equation}\label{sigmadef}
\sigma(s)=\frac1{\beta_0}\theta(\Lambda^2+s)\theta(-s)+\frac1{\pi\beta_0}
  \theta(s)\arccos\left(\frac{\ln(s/\Lambda^2)}{\sqrt{\pi^2
  +\ln^2(s/\Lambda^2})}\right).
\end{equation}
This formal result can be reformulated as integration over the spectrum
$\sigma(s)$ using Cauchy's theorem. Indeed, from Eq.~(\ref{resumcorr}) one
readily deduces the singularity of the resummed polarization function with the
discontinuity
\begin{equation}
{\rm Disc\,}\Pi(s)=\frac{2\pi i}{\beta_0}\Bigg\{\theta(\Lambda^2+s)
  \theta(-s)+\frac1\pi\theta(s)\arccos
  \left(\frac{\ln(s/\Lambda^2)}{\sqrt{\pi^2+\ln^2(s/\Lambda^2})}\right)\Bigg\}
\end{equation} 
which coincides with $\sigma(s)$ in Eq.~(\ref{sigmadef}) and, thereby, simply
represents the spectrum.

\begin{figure}
\begin{center}
\epsfig{figure=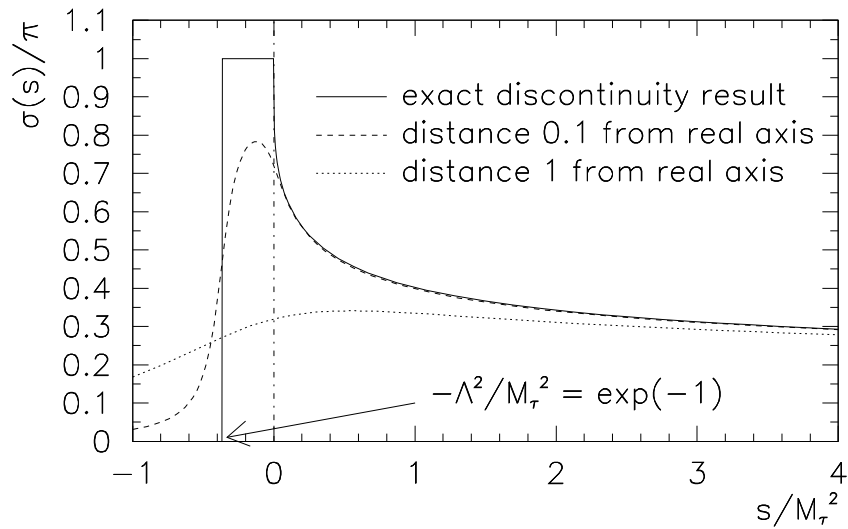,scale=0.7}
\caption{\label{fig8}The spectral function $\sigma(s)$ as a function of $s$.
  Shown is the exact result for the discontinuity (full line) as well as
  results for non-closed circle integrals approaching the axis along a full
  circle from both sides, see Fig.~\ref{fig2}. The numerical values for the
  distances orthogonal to the real axis towards the positive and negative
  imaginary semi-plane in $s$ are given in units of $M_\tau^2$.}
\end{center}
\end{figure}

This spectral function $\sigma(s)$ is shown in Fig.~\ref{fig8}. The part of
the spectrum on the positive real axis is an analytic continuation of the
renormalization group improved correlator function $\Pi(Q^2)$ to the
cut~\cite{picorr,Pivtau,rad,kim,shirkov}. It can be conveniently written in
the form
\begin{equation}\label{arccostg}
\arccos\left(\frac{\ln(s/\Lambda^2)}{\sqrt{\pi^2+\ln^2(s/\Lambda^2)}}\right)
  =\arcsin\left(\frac{\pi}{\sqrt{\pi^2+\ln^2(s/\Lambda^2)}}\right)
  =\arctan\left(\frac{\pi}{\ln(s/\Lambda^2)}\right)
\end{equation}
with the appropiate branches of the relevant functions. Introducing
\begin{equation}\label{assimp}
\alpha(s)=\frac\pi{\beta_0\ln(s/\Lambda^2)}
\end{equation}
one can write
\begin{equation}
\arctan\left(\frac\pi{\ln(s/\Lambda^2)}\right)=\arctan(\beta_0\alpha(s))
\end{equation}

Taking the value of the spectral density at $M_\tau^2$ as initial value,
\begin{equation}
\sigma(M_\tau^2)=\frac1{\pi\beta_0}\arctan(\beta_0\alpha_\tau)
  =\frac{\alpha_\tau}\pi+O(\alpha^2).
\end{equation}
we can introduce the continuum part of the spectral density,
\begin{equation}\label{sigc}
\sigma_c(s)=\frac1{\pi\beta_0}\arctan(\beta_0\alpha(s)).
\end{equation}
The differential equation determining the continuum part $\sigma_c(s)$
through its initial value $\sigma(M_\tau^2)$ can be constructed by
differentiating Eq.~(\ref{sigc}) with respect to $s$,
\begin{equation}
s\frac{d\sigma_c(s)}{ds}
  =-\beta_0\pfrac{\alpha(s)}\pi^2\frac1{1+\beta_0^2\alpha(s)^2}
\end{equation}
with $\alpha(s)$ taken from Eq.~(\ref{assimp}). By inverting
Eq.~(\ref{sigc}) we have
\begin{equation}
\beta_0\alpha(s)=\tan(\pi\beta_0\sigma_c(s))\quad\mbox{for\ }s>0
\end{equation}
and therefore obtain
\begin{equation}\label{sigmacev}
s\frac{d}{ds}\sigma_c(s)=-\frac1{\pi^2\beta_0}\sin^2(\pi\beta_0\sigma_c(s))
  \quad\mbox{for\ }s>0.
\end{equation}
This equation can indeed be considered as an evolution equation for the
spectral density $\sigma_c(s)$ determining $\sigma_c(s)$ through its initial
value $\sigma(M_\tau^2)$. Therefore, one can simply introduce an effective
charge
\begin{equation}
\alpha_M(s)=\pi\sigma_c(s)\quad\mbox{or}\quad
  a_M(s)=\frac{\alpha_M(s)}\pi=\sigma_c(s)
\end{equation}
(see Eq.~(\ref{rhobyeff})). The evolution equation for $a_M(s)$ is then given
by
\begin{equation}\label{almev}
s\frac{da_M(s)}{ds}=-\frac1{\pi^2\beta_0}\sin^2(\pi\beta_0a_M(s)).
\end{equation}
Therefore, we now define the coupling as the value of the spectral density on
the cut far from the IR region. This is a perturbation theory definition. The
evolution of this coupling, however, is calculated by taking into account the
analytic continuation. Then it has an IR fixed point with the coupling value
\begin{equation}\label{IRfix}
a_M(0)=1/\beta_0
\end{equation}
(we assume that all transformations are made for small $a_M$ so that this is
the first IR point).

Starting from finite-order perturbation theory one therefore can find the
all-order result because of analytic continuation and running in the IR domain
-- it happens to be finite (no poles for the spectrum at small values of $s$).
Therefore, the function $\sigma_c(s)$ can be used to generate moments by
integration on the cut. Note that in this approach the part of the spectrum
at negative $s$ is lost and cannot be recovered from finite-order perturbation
theory.

Note, however, that the value at zero mentioned in Eq.~(\ref{IRfix}) is not a
universal quantity. If Adler's function starts with a proper power of the
coupling constant as it is the case for gluonic observables, for instance,
this picture will change. For instance, we take
\begin{equation}
D(Q^2)=\pfrac{\alpha_E(Q^2)}\pi^2.
\end{equation}
Then the corresponding spectral density in leading order $\beta$-function
approximation reads
\begin{equation}
\rho(s)=\frac1{\beta_0^2}\ \frac1{\ln^2(s/\Lambda^2)+\pi^2}
  =\frac{\alpha^2(s)}{\pi^2(1+\beta_0^2\alpha^2(s))}.
\end{equation}
As in previous cases, we can define an effective coupling in the Euclidean
domain,
\begin{equation}\label{effM2}
\bar a_M(s)=\frac{\alpha(s)}{\pi\sqrt{1+\beta_0^2\alpha^2(s)}}.
\end{equation}
For this effective coupling we indeed have (cf Eq.~(\ref{IRfix}))
\begin{equation}
\bar a_M(0)=0.
\end{equation}
The $\beta$-function for the effective coupling obtained from
Eq.~(\ref{effM2}),
\begin{equation}
\bar\beta_M(\bar a_M)=-\beta_0\bar a_M^2\sqrt{1-(\pi\beta_0\bar a_M)^2}
  =-\beta_0\bar a_M^2\left(1-\frac12(\pi\beta_0\bar a_M)^2+O(\bar a_M^4)\right)
\end{equation}
differs from the $\beta$-function of the effective coupling $a_M$ (c.f.\
Eq.~(\ref{almev}))
\begin{equation}
\beta_M(a_M)=-\frac1{\pi^2\beta_0}\sin\left(\pi\beta_0a_M\right)
  =-\beta_0a_M^2\left(1-\frac13(\pi\beta_0a_M)^2+O(a_M^4)\right)
\end{equation}
at next-to-leading order.

\begin{figure}
\begin{center}
\epsfig{figure=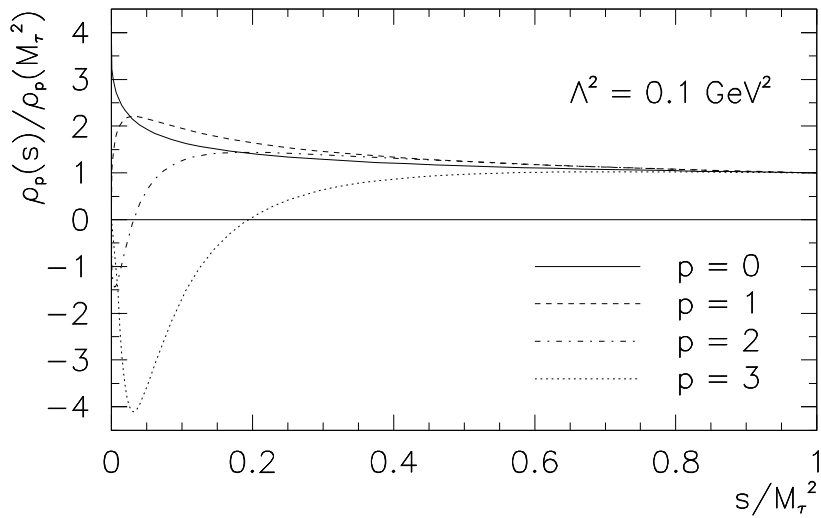,scale=0.7}
\caption{\label{fig9}The spectral function $\rho_p(s)$ as a function of $s$,
  given by Eq.~(\ref{rhop}). All spectral functions for $p>0$ are going to
  zero for $s\rightarrow 0$, for $p\ge 2$ they have nontrivial zeros and
  a fluctuating behaviour near the origin.}
\end{center}
\end{figure}

The consideration can be generalized to higher powers of the coupling in the
Euclidean domain. For
\begin{equation}
D_p(Q^2)=\pfrac{\alpha_E(Q^2)}\pi^{p+1}
\end{equation}
we obtain
\begin{equation}\label{rhop}
\rho_p(s)=\frac1{\pi\beta_0}\ \frac{\sin(p\phi(s))}{pr^p(s)}
\end{equation}
where
\begin{equation}
r(s)=\beta_0\sqrt{\ln^2(s/\Lambda^2)+\pi^2},\qquad
\tan\phi(s)=\frac{\pi}{\ln(s/\Lambda^2)}=\beta_0\alpha(s).
\end{equation}
For $p=0$ we retain Eq.~(\ref{sigc}),
\begin{equation}
\rho_0(s)=\frac1{\pi\beta_0}\phi(s)
  =\frac1{\pi\beta_0}\arctan(\beta_0\alpha(s)).
\end{equation}
But for large values of $p$ the function $\rho_p(s)$ starts to fluctuate when
starting at $\rho(M_\tau^2)$ and approaching $s=0$ where it finally reaches
zero. Because if this behaviour, shown in Fig.~\ref{fig9} in some detail for
the first four values of $p$, $\rho_p(s)$ fails to be interpreted as a
(positive) spectral density in a region which enlarges with increasing values
of $p$. 

\begin{figure}
\begin{center}
\epsfig{figure=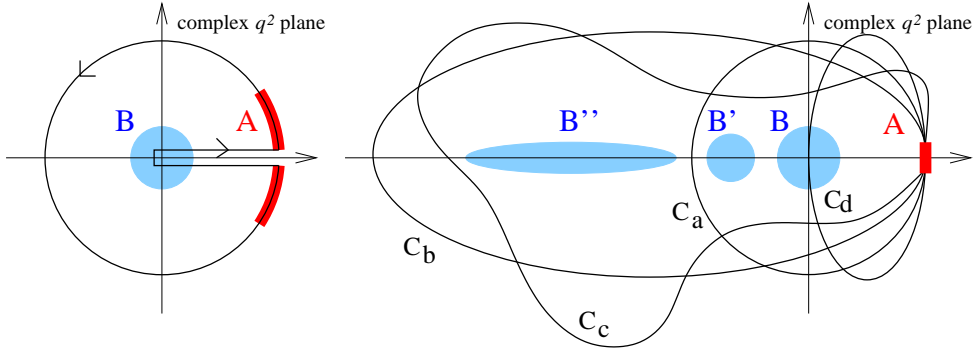, scale=0.7}
\caption{\label{figa}Contours in the complex plane with a fixation in the
  region A, taking into account possible occurrences of singularities. The
  figure part on the left hand side shows the standard circle path which
  circumvents the singular region B while the occurence of other singularity
  regions as discussed in the text (regions B$'$ and B$''$) may lead to
  different possibilities for choosing a path (C$_{\rm a}$, C$_{\rm b}$,
  C$_{\rm c}$). The path C$_{\rm d}$ crosses the singularity region and,
  therefore, cannot be used from the perturbation theory point of view.}
\end{center}
\end{figure}

We conclude that the two recipes of resummation (on the contour and on the
positive semi-axis for $s$) are different: they differ by the integral over
the negative real semi-axis for $s$ or, more generally, by the integral
passing through the infrared region. This point is worth discussing in more
detail. In the contour formulation it is not essential what particular
point-by-point behaviour in the IR region exists. For the analytically
continued correlator this is not important unless the contour crosses a
nonanalytic region. Whatever singularities exist in the IR region (Regions B,
B', or B'' in Fig.~\ref{figa}), the contour includes them. This is the
definition of resummation on the contour which is explicitly perturbative.

For the formulation on the cut the extrapolation of the running of the
coupling constant to the IR region is crucial because it provides an
interpretation of the spectral density in the IR region. One can formally
write 
\begin{equation}\label{Irdirect}
\int_0^{M_\tau^2}\alpha(s)ds
\end{equation}
and try to interpret this integral. The naive substitution of the expression
in Eq.~(\ref{assimp}) leads to a singularity which is known as Landau pole.
Still, further formal manipulations
readily give ($t=\beta_0\alpha(M_\tau^2)\ln(M_\tau^2/s)$)
\begin{equation}\label{IrBorel}
\int_0^{M_\tau^2}\alpha(s)ds=\int_0^{M_\tau^2}
  \frac{\alpha(M_\tau^2)ds}{1+\beta_0\alpha(M_\tau^2)\ln(s/M_\tau^2)}
  =\frac{M_\tau^2}{\beta_0}\int_0^{\infty}
  \frac{e^{-t/\beta_0\alpha(M_\tau^2)}dt}{1-t}.
\end{equation}
Both equations are not well-defined from the beginning, the second form being
a Borel representation. The same problem can be reformulated as a divergence
of the asymptotic series. Indeed, by expanding the expression for the running
coupling under the integration sign in a perturbation theory series one has
\begin{equation}\label{serIR}
\int_0^{M_\tau^2}\alpha(s)ds=\sum_nn!\pfrac{\beta_0\alpha(M_\tau^2)}\pi^n.
\end{equation}
The summation of the series in Eq.~(\ref{serIR}) is equivalent to the
interpretation of the integral, independent on what form is chosen for it,
so either Eq.~(\ref{Irdirect}) or Eq.~(\ref{IrBorel}). Therefore, an
integrable behaviour of the coupling constant at small $s$ (whatever the
definition of this quantity is) can solve the problem of divergence of the
integral or the summation of the asymptotic series. But this solution is
strongly model dependent because the extrapolation of the evolution into the
IR region is essentially arbitrary and higher order corrections are important. 
The explicit form of the extrapolation in
Eq.~(\ref{sigmadef}) gives an extrapolation motivated by analytic
continuation. It can also be considered as a special change of the
renormalization scheme~\cite{renRS}. Indeed, for the coupling $a_M$ with the
evolution given in Eq.~(\ref{almev}) one obtains an IR fixed point. In any
case the result depends on the IR region which is strongly nonperturbative.

As it has become popular as being used as an extrapolation, for the analytic
continuation the moments are defined simply by
\begin{equation}\label{ancontmom}
m_{00}^{\rm anal}=\int_0^{M_\tau^2}\frac{\sigma_c(s)ds}{M_\tau^2},
\end{equation}
excluding totally the negative part of the spectrum that emerges in the exact
treatment of quantities originally defined in the Euclidean domain (see
Eq.~(\ref{m00result})). Using Eq.~(\ref{m00further}) one can relate these
moments to the moments on the contour by
\begin{equation}\label{ancontvsmymom}
m_{00}^{\rm anal}=-\frac1{\beta_0}e^{-\pi/\beta_0\alpha_\tau}+m_{00}.
\end{equation}

The contour definition for the moments is a strictly perturbative one and in
this particular case simply gives the analytic function of $\alpha_\tau$ at
the origin $\alpha_\tau=0$, i.e.\ $m_{00}(\alpha_\tau)$ is analytic at small
$\alpha_\tau$. Then the decomposition in Eq.~(\ref{ancontvsmymom}) is at least
meaningful, the right hand side does contain terms of different analytic
structure. We can, of course, reproduce the contour moments by summing on the
cut (which gives $m_{00}^{\rm anal}$) and adding a pole (giving the
``nonperturbative'' term in Eq.~(\ref{ancontvsmymom})) which is an allowed
extrapolation as well (because the IR region behaviour cannot be found in
perturbation theory). But it is difficult to imagine that one could suggest
such a treatment of the situation just as it is, if not derived naturally from
the contour definition. Note that this is completely analogous to the
situation with Coulomb resummation: by summing on the cut $4m^2<s<\infty$ one
can never restore the pole contributions generated by the bound states. Only
an explicit analytic continuation can reveal the whole analytic structure of
the correlator. The result of Eq.~(\ref{ancontvsmymom}) can readily be
generalized to any power of $\alpha_\tau$ as a basic quantity which is
important for correlators of hadronic currents with nonvanishing anomalous
dimensions. 

Note that the problem of a particular choice for the extrapolation of the
evolution into the IR region cannot be uniquely solved from the point of view
of the perturbation theory because the IR region is nonperturbative and there
is no tool to choose a ``correct'' extrapolation in cases when there is no
explicit analytic structure known. Therefore, the extrapolation is rather
arbitrary for basically all cases except the case of two-point correlators.
Still, a particular choice for the extrapolation can be important numerically
for the theoretical description of the available high precision data. In
$\tau$ decays the special role of contour moments is related to perturbation
theory -- they are really computable in perturbation theory without any
additional hypotheses. For the $s^l$-moments the problem is less important
numerically while for the $(M_\tau^2-s)^k$-moments it is crucial (see
Table~\ref{tab5}). This is obvious because the $(M_\tau^2-s)^k$-moments are
saturated at small $s$ and the particular way of treating this region is
important. In order to discuss this issue, we now leave the special case
$k=l=0$ and turn to the description of modified moments, discussing the
differences. 

\subsection{The moments $m_{0l}$}
In general, for any given moment $m_{0l}$ one finds
\begin{eqnarray}\label{mom0lcirc}
m_{0l}&=&\frac1{\pi\beta_0}\Bigg\{\phi+\sum_{j=1}^{n-1}(j-1)!
  \pfrac{\beta_0\alpha_\tau}{(l+1)\pi r}^j\sin(j\phi)\,+\nonumber\\&&-(l+1)
  (-1)^{l+1}\frac{(n-1)!}2\pfrac{\beta_0\alpha}{(l+1)\pi}^n\int_{-\pi}^\pi
  \frac{e^{i(l+1)\varphi}d\varphi}{(1+i\beta_0\alpha_\tau\varphi/\pi)^n}
  \Bigg\}.
\end{eqnarray}
The representation obtained by using integration by parts shows an improvement
in the convergence for moments at large $l$ equivalent to the replacement
$\alpha_\tau\to\alpha_\tau/(l+1)$. This coincides with conclusions drawn from
finite-order perturbation theory. For arbitrary $l$ one can also find a
representation containing an explicit nonperturbative term,
\begin{eqnarray}
m_{0l}&=&\frac1{\pi\beta_0}\Bigg\{-\pi(-1)^{l+1}
  e^{-(l+1)\pi/\beta_0\alpha_\tau}+\phi+\sum_{j=1}^{n-1}(j-1)!
  \pfrac{\beta_0\alpha_\tau}{(l+1)\pi r}^j\sin(j\phi)\nonumber\\&&
  +(l+1)(-1)^{l+1}\frac{(n-1)!}{2}\pfrac{\beta_0\alpha}{(l+1)\pi}^n
  \left(\int_{-\infty}^{-\pi}+\int_\pi^\infty\right)\frac{e^{i(l+1)\varphi}
  d\varphi}{(1+i\beta_0\alpha_\tau\varphi/\pi)^n}\Bigg\}.
\end{eqnarray}
These formulae give a direct generalization to large $l$ moments and
can be treated perturbatively.

\subsection{The moments $m_{k0}$}
The moments for $l=0$ corresponding to the $\alpha_\tau$ dependent part of the
correlator in the leading order are given by
\begin{equation}
m_{k0}=-\frac{k+1}{2\pi\beta_0}\int_{-\pi}^\pi
  (1+e^{i\varphi})^k\ln(1+i\beta_0\alpha_\tau\varphi/\pi)e^{i\varphi}d\varphi.
\end{equation}
These moments can also be expanded in a convergent series in $\alpha_\tau$
for small values of $\alpha_\tau$, recovering the results from finite-order
perturbation theory.

The usual arguments given for the preferred use of these moments from the
theoretical point of view is that they suppress contributions close to the
real axis where OPE is not applicable (region A in Fig.~\ref{figa})). They do
suppress these contributions because of $e^{i\varphi}\sim -1$ in this region.
One can also obtain this result by looking at the integral measure for the
moments $m_{k0}$,
\begin{equation}
(1+e^{i\varphi})^ke^{i\varphi}d\varphi
  =(2\cos(\varphi/2))^ke^{ik\varphi/2}e^{i\varphi}d\varphi
\end{equation}
where the suppression is given by the factor $(1+e^{i\varphi})^k$. The
suppression of oscillations starts at larger values of the integration
parameter $\varphi$ than for the moments $m_{0l}$ but there is a suppression
by the power of $\cos(\varphi/2)$, $\cos(\varphi/2)=0$ at $\varphi=\pm\pi$.
However, one can see that these moments are almost insensitive to the value of
$\alpha_\tau$ for large $k$ (or, more precisely, to the perturbative
structure). It is clear that these functions determine integrals which
really have no relation to perturbative running because such a running is
completely suppressed. This makes an asymptotic expansion in $\alpha_\tau$
almost useless (we could not find any efficient asymptotic expansion in
$\alpha_\tau$ for this moments at large values of $k$). However, it is
straightforward to write down a nonperturbative representation for these
moments at large values of $k$. A representation in terms of integral over the
spectrum is particularly convenient. One has
\begin{equation}
m_{k0}=-\frac{k+1}{2\pi\beta_0}\int_{-\pi}^\pi
  (1+e^{i\varphi})^k\ln(1+i\beta_0\alpha_\tau\varphi/\pi)e^{i\varphi}d\varphi
  =\int_{-\Lambda^2}^{M_\tau^2}\left(1-\frac{s}{M_\tau^2}\right)^k
  \frac{\sigma(s)ds}{M_\tau^2}.
\end{equation}
The part of the spectrum on the negative real axis is given by a constant. For
large values of $k$ the integral is saturated at the left boundary of the
integration region and reads
\begin{equation}
m_{k0}=\frac1{\beta_0}\left(\left(1+\frac{\Lambda^2}{M_\tau^2}\right)^{k+1}
  -1\right)+(k+1)\int_0^{M_\tau^2}\left(1-\frac{s}{M_\tau^2}\right)^k
  \frac{\sigma_c(s)ds}{M_\tau^2}.
\end{equation}
The contribution of the integral is suppressed because the spectrum 
$\sigma_c(s)$ is a smooth and finite function leading to a finite result
for the integral.

Another way to look at the problem is given by integrating by parts once.
One has
\begin{equation}
m_{k0}=\frac{\alpha_\tau}{2\pi^2}\int_{-\pi}^\pi
  \frac{(1+e^{i\varphi})^{k+1}d\varphi}{1+i\beta_0\alpha_\tau\varphi/\pi}.
\end{equation}
In this case the boundary terms vanish. One obtains
\begin{equation}\label{finalkmom}
m_{k0}=\frac1{\beta_0}\left(1+\frac{\Lambda^2}{M_\tau^2}\right)^{k+1}
  -\frac1{\beta_0}\int_0^{M_\tau^2}\left(1-\frac{s}{M_\tau^2}\right)^{k+1}
  \frac1{\pi^2+\ln^2(s/\Lambda^2)}\frac{ds}{s}.
\end{equation}
Here it is easy to estimate the contribution of the integral as well. One has
\begin{equation}
\int_0^{M_\tau^2}\left(1-\frac{s}{M_\tau^2}\right)^{k+1}
  \frac1{\pi^2+\ln^2(s/\Lambda^2)}\frac{ds}s
  <\int_0^{M_\tau^2}\frac1{\pi^2+\ln^2(s/\Lambda^2)}\frac{ds}s<1.
\end{equation}
Note that the whole function $m_{k0}$ is still represented by a convergent
series for small values of $\alpha_\tau$.

\begin{table}\begin{center}
\begin{tabular}{|l|rrrrrrrr|}\hline
$l=3$&$-0.000$&$-0.000$&$-0.000$&$-0.000$&$-0.000$&$-0.001$&$-0.002$&$-0.003$\\
$l=2$&$+0.000$&$+0.002$&$+0.004$&$+0.008$&$+0.013$&$+0.020$&$+0.030$&$+0.042$\\
$l=1$&$-0.003$&$-0.009$&$-0.018$&$-0.032$&$-0.052$&$-0.078$&$-0.113$&$-0.160$\\
$l=0$&$+0.186$&$+0.305$&$+0.396$&$+0.469$&$+0.530$&$+0.582$&$+0.626$&$+0.665$\\
\hline
&$k=0$&$k=1$&$k=2$&$k=3$&$k=4$&$k=5$&$k=6$&$k=7$\\\hline
\end{tabular}
\caption{\label{tab5}Contribution of the integral taken over the interval from
  $-\Lambda^2$ to $+\Lambda^2$ for the moments $m_{kl}$ in Eq.~(\ref{momklsig})
  relative to the integral taken over to the whole integration range.}
\end{center}\end{table}

In general we determine the contribution of the nonperturbative region
determined by the scale $\Lambda^2$ to the modified moments
\begin{equation}\label{momklsig}
m_{kl}=\int_{-\Lambda^2}^{M_\tau^2}\left(1-\frac{s}{M_\tau^2}\right)^k
  \pfrac{s}{M_\tau^2}^l\frac{\sigma(s)ds}{M_\tau^2}.
\end{equation}
in order to consider the enhancement of this region in relation to the
moment as a whole. In Table~\ref{tab5} we give values for the contribution
of the interval range $[-\Lambda^2,\Lambda^2]$ to the contribution for the
whole range $[-\Lambda^2,M_\tau^2]$ for different values of $k$ and $l$,
taking $\Lambda^2=0.2\GeV^2$. A strong enhancement can be observed especially
for the case $l=0$ and large values for $k$. One observes that the first term
is larger than the second one already for $k>3$. Though weakened, the same
tendency can be seen in Table~\ref{tab6} where we considered the contribution
of the interval range $[-\Lambda^2,0]$ relative to the whole range.

This observation also suggests the question of sensitivity of the moments to
the parameters describing  the strength of strong interactions, namely the
values of $a_M^\tau$ and  $\Lambda$. As expected, the $s^l$-moments are
sensitive to the spectrum at large energies, i.e.\ $\sigma(M_\tau^2)$ which is
almost equal to $a_M^\tau$. The large $k$ moments are directly sensitive to
$\Lambda$ as one can see from Eq.~(\ref{finalkmom}). However, perturbation
theory is insufficient for these moments, and thus the theoretical formulae
for large $k$ moments like Eq.~(\ref{finalkmom}) cannot be confronted with
experiment and should be supplemented by a quantitative hadronization
mechanism.

\begin{table}\begin{center}
\begin{tabular}{|l|rrrrrrrr|}\hline
$l=3$&$-0.000$&$-0.000$&$-0.000$&$-0.000$&$-0.001$&$-0.001$&$-0.002$&$-0.004$\\
$l=2$&$+0.000$&$+0.001$&$+0.003$&$+0.005$&$+0.009$&$+0.014$&$+0.022$&$+0.031$\\
$l=1$&$-0.006$&$-0.018$&$-0.036$&$-0.061$&$-0.093$&$-0.134$&$-0.187$&$-0.256$\\
$l=0$&$+0.117$&$+0.195$&$+0.256$&$+0.308$&$+0.353$&$+0.392$&$+0.428$&$+0.460$\\
\hline
&$k=0$&$k=1$&$k=2$&$k=3$&$k=4$&$k=5$&$k=6$&$k=7$\\\hline
\end{tabular}
\caption{\label{tab6}Contribution of the integral taken over the interval from
  $-\Lambda^2$ to $0$ for the moments $m_{kl}$ in Eq.~(\ref{momklsig})
  relative to the integral taken over to the whole integration range.}
\end{center}\end{table}

\subsection{Higher orders of running}
The formulae obtained for the polarization function can readily be generalized
to higher orders of the $\beta$-function. For Adler's function one has
\begin{equation}
D(Q^2)=1+a_E(Q^2).
\end{equation}
The polarization function is then determined by the equation
\begin{equation}
-Q^2\frac{d}{dQ^2}\Pi(Q^2)=1+a_E(Q^2).
\end{equation}
A simple way to calculate $\Pi(Q^2)$ is to use the ansatz
\begin{equation}
\Pi(Q^2)=\ln\pfrac{\mu^2}{Q^2}+\ln f\left(a_E(Q^2)\right).
\end{equation}
One then obtains a differential equation for $f(a_E)$ which reads
\begin{equation}
-Q^2\frac{d}{dQ^2}\Pi(Q^2)=1-\frac1{f(a_E(Q^2))}Q^2\frac{d}{dQ^2}
  f\left(a_E(Q^2)\right)=1-\frac{f'(a_E(Q^2))}{f(a_E(Q^2))}Q^2
  \frac{da_E(Q^2)}{dQ^2}.
\end{equation}
Using the renormalization group equation
\begin{equation}
Q^2\frac{da_E(Q^2)}{dQ^2}=\beta\left(a_E(Q^2)\right)
\end{equation}
one finds
\begin{equation}
f'\left(a_E(Q^2)\right)\beta\left(a_E(Q^2)\right)
  =-f\left(a_E(Q^2)\right)a_E(Q^2).
\end{equation}
This differential equation can be solved in quadrature,
\begin{equation}
\ln f=-\int\frac{a_Eda_E}{\beta(a_E)}.
\end{equation}
The solution for $\Pi(Q^2)$ is therefore given by the trajectory
\begin{equation}\label{traject}
\Pi(Q^2)=-\int\frac{dQ^2}{Q^2}D(Q^2)
  =\ln\pfrac{\mu^2}{Q^2}-\int\frac{a_Eda_E}{\beta(a_E)}.
\end{equation}
Eq.~(\ref{traject}) can be used for any $\beta$-function (with the
restriction that no singularity should occur). For $\beta(a)=-\beta_0a^2$ we
obtain the former result,
\begin{equation}
\beta(a)=-\beta_0a^2\quad\Rightarrow\quad\ln f=\frac1{\beta_0}
  \ln\left(a_E(Q^2)\right).
\end{equation}
For $\beta(a)=-\beta_0a^2-\beta_1a^3$ we explicitly obtain
\begin{equation}
\Pi(Q^2)=\ln\pfrac{\mu^2}{Q^2}+\frac1{\beta_0}
  \ln\pfrac{\beta_0a_E(Q^2)}{\beta_0+\beta_1a_E(Q^2)}.
\end{equation}
The generalization to any polynomial function that may occur in perturbation
theory is straightforward. 

As a last note we comment on the Landau singularity which occurs at the
leading order of the running. For two-point correlators the problem is
completely solved by the contour integration to any order of the
$\beta$-function. We stress that the pole is inside the circle (see the
discussion in Ref.~\cite{pivsuppl}). One cannot deform the contour to be close
to the origin because this region is completely nonperturbative and should
therefore be avoided. This situation is completely analogous to the situation
with Coulombic poles~\cite{coulK} or with a spectrum below threshold for
a gluonic operator induced by a heavy quark loop~\cite{last}. Any type of
singularity is avoided by moving the contour far from the origin and keeping
the singularity inside, thereby including also the integral of the spectrum.

\section{Conclusion}
We have shown that the direct $s^l$-moments are perturbative and can be
reliably calculated within operator product expansion (OPE). The mixed moments
with the weight functions $(1-s/M_\tau^2)^k$ are strongly nonperturbative.
This is natural and is expected in QCD with asymptotic freedom. We have
analyzed finite-order perturbation theory, power corrections and resummation.
The results agree with each other and the picture is completely consistent. An
asymptotic analysis (large values for $l$ or $k$) definitely disfavours the
large $k$ moments despite arguments that they suppress the non-OPE region
along the contour -- these arguments are not valid. In practical applications
to $\tau$ decays and the determination of $m_s$ the second mixed moment with
weight function $(1-s/M_\tau^2)^2$ is sometimes used. Along with the initial
phase space weight $(1-s/M_\tau^2)^2$ (see Eq.~(\ref{int})) the total results
gives a moment with the weight $(1-s/M_\tau^2)^4$ which is marginal in the
sense that it enhances nonperturbative contributions. Our consideration show
that this moment is really on the border of what is still theoretically
computable. The theoretical error becomes larger than one would expect for
such an analysis. Note that an estimate of the systematic theoretical error,
i.e.\ the error due to the functional dependence like perturbation series
truncation, is complicated as compared to a simple statistical error of
theoretical formulae due to uncertainties in the parameters used. The
theoretical error for the moment with weight $(1-s/M_\tau^2)^4$ appearing in
the phenomenological analysis is big enough to be concerned. The point is
whether we should worry about the size of this error in comparison with the
gain due to the improvement in the accuracy of the experimental moments or to
trade this moment for a more reliable theoretical expression confronted with a
less precise experimental moment. This still remains to be seen.

\subsection*{Acknowledgements}
We topics of the present paper were discussed with many people at different
times. We acknowledge discussions and correspondence with M.~Beneke,
K.~Chetyrkin, A.~Kataev, N.~Krasnikov, J.~K\"uhn, K.~Maltman, S.~Narison, 
K.~Schilcher, D.~Shirkov, N.~Uraltsev, and V.~Zakharov. A.A.~Pivovarov is
indebted to V.~Matveev for support and interest in the work.

The present work is supported in part by the Russian Fund for Basic Research
under contracts 99-01-00091 and 01-02-16171. A.A.~Pivovarov is Alexander von
Humboldt fellow. S.~Groote acknowledges a grant given by the Deutsche
Forschungsgemeinschaft, Germany.

\end{document}